\documentstyle[aps,prd,floats,graphicx,eqsecnum]{revtex}

\def\be{\begin{equation}}
\def\ee{\end{equation}}
\def\bea{\begin{eqnarray}}
\def\eea{\end{eqnarray}}
\def\ab{a_b}
\def\nb{n_b}
\def\A{A}
\def\Ay{{A_y}}
\def\Ayy{{A_{yy}}}
\def\S{S}
\def\Sy{S_y}
\def\R{{\cal{R}}}
\def\B{{\cal{B}}}
\def\s{{\cal{S}}}
\def\T{T}
\def\t{\tau}
\def\ty{\tau_y}
\def\ei{\hat e_i}
\def\eij{\left( \hat{e}^i_{~,j}+\hat{e}_{j,}^{~i}\right)}

\def\sigy{{\sigma_y}}
\def\Xfive{{}^{(5)}\delta G_T}
\def\Yfive{{}^{(5)}\delta G_{TF}}
\def\XE{\delta E_T}
\def\YE{\delta E_{TF}}
\def\Xfour{{}^{(4)}\delta {G}_T}
\def\Yfour{{}^{(4)}\delta {G}_{TF}}
\def\dpvec{\delta p^{\rm{(vector)}}}
\def\dpivec{\delta \pi^{\rm{(vector)}}}
\def\dpitens{\delta \pi^{i~\rm{(tensor)}}_j}
\def\lAyy{{\tilde{A}_{yy}}}
\def\lAy{{\tilde{A}_{y}}}
\def\lR{{\tilde{\cal{R}}}}
\def\lA{{\tilde{A}}}

\newcommand{\square}{\kern1pt\vbox{\hrule height  1.2pt\hbox{\vrule
width 1.2pt\hskip 3pt \vbox{\vskip 6pt}\hskip  3pt\vrule width
0.6pt}\hrule height 0.6pt}\kern1pt}
\begin{document}
\draft

\preprint{}

\title{Cosmological perturbations in the bulk and on the brane}

\author{Helen A.~Bridgman, Karim A.~Malik and David Wands}

\address{Relativity and Cosmology Group, School of Computer Science
and Mathematics,\\University of Portsmouth, Portsmouth~PO1~2EG,
United Kingdom}

\date{\today}
\maketitle

%
%

\begin{abstract}
  We study cosmological perturbations in a brane-world scenario where
  the matter fields live on a four-dimensional brane and gravity
  propagates in the five-dimensional bulk. We present the equations of
  motion in an arbitrary gauge for metric perturbations in the bulk
  and matter perturbations on the brane. Gauge-invariant
  perturbations are then constructed corresponding to perturbations
  in longitudinal and Gaussian normal gauges.
  Longitudinal gauge metric perturbations may be directly derived from
  three master variables (separately describing scalar, vector and tensor
  metric perturbations) which obey five-dimensional wave-equations.
  Gaussian normal gauge perturbations are directly related to the
  induced metric perturbations on the brane with the additional
  bulk degrees of freedom interpreted as an effective Weyl
  energy-momentum tensor on the brane.
  We construct gauge-invariant perturbations describing the effective
  density, momentum and pressures of this Weyl fluid at the brane and
  throughout the bulk.
  We show that there exist gauge-invariant
  curvature perturbations on the brane and in the bulk that are
  conserved on large-scales when three-dimensional spatial gradients
  are negligible.
\end{abstract}
\vskip 1pc \pacs{98.80.Cq \ \ 04.50.+h
\hfill PU-RCG-01/24 \ \ astro-ph/0107245 \ \ 
Phys.~Rev.~D~{\bf{65}}, 043502   \copyright 2002 APS}
%

\section{Introduction}

The brane-world scenario offers a different approach to dimensional
reduction from the traditional Kaluza-Klein model.  Ordinary matter
fields are assumed to be confined to a lower-dimensional brane in a
higher-dimensional spacetime rather than being associated with bulk
degrees of freedom. The gravity that the brane-bound observer feels is
only the projected gravity of a higher-dimensional gravitational
field~\cite{BDL,SMS,MW,MB,BCMU}. In this paper we will consider 
models where the bulk is described by Einstein gravity in which case we
require a positive brane tension for the induced gravity to have a
positive gravitational constant at low energies, and a negative
cosmological constant in the bulk to have a vanishing induced
cosmological constant on the brane. Randall and Sundrum~\cite{RS2}
demonstrated that in such a model it is possible to have a
four-dimensional Minkowski brane-world in five-dimensional anti-de
Sitter spacetime (AdS$_5$) where gravity is effectively
four-dimensional at low-energies even if the bulk is non-compact.

Our aim in this paper is to describe arbitrary linear perturbations
about 4D Friedmann-Robertson-Walker (FRW) cosmologies embedded in 5D
Schwarzschild-Anti-de Sitter bulk~\cite{BDEL,MSM,BCG} and understand
the effect of bulk metric perturbations on the brane.  Modelling the
generation and evolution of cosmological perturbations offers a
means by which current brane-world scenarios can be
empirically tested. Although considerable effort has already been
devoted to this subject
\cite{GS,Mukoh,Roy,Langlois,HHR,LMW,cvdb,Koyama,BMW,Deruelle,Sasaki,LMSW},
the results known thus far are limited to special cases. The only
cosmological background in which the five-dimensional perturbation
equations are separable corresponds to a de Sitter (or Minkowski)
brane \cite{Kaloper,RS2,GS,HHR,LMW,BMW}. In more general (and more
realistic) cosmological settings current results are restricted to
either small (3D) scales \cite{Koyama} where the cosmological
expansion may be neglected, or large scales \cite{LMSW} where gradient
terms are neglected, and even then our understanding is
incomplete~\cite{Roy}.

We will assume that our 4-dimensional world is described by a domain
wall (3-brane) $(M,g_{\mu\nu})$ in the 5-dimensional spacetime
$({\cal M},{}^{(5)}g_{AB})$.
We will denote the vector unit normal to $M$ by $n^A$ and the induced
metric on $M$ by
\begin{equation}
g_{AB} = {}^{(5)}g_{AB} - n_{A}n_{B}
\end{equation}
%
The effective action in the 5-dimensional spacetime is 
\begin{eqnarray}
S= \int_{{\cal M}} d^5X \sqrt{-{}^{(5)}g} \left[ {1 \over 2
\kappa_5^2} \left( {}^{(5)}R  - 2\Lambda_5 \right) \right]
 + \int_M d^4 x \sqrt{-g}
\left[ L_{\rm matter} -  \lambda \right] ,
\end{eqnarray}
where $x^\mu$ are the 4-dimensional coordinates induced on the
brane world.
%
The 5-dimensional Einstein equations, obtained by minimising the
action with respect to variations in the bulk metric, are
\begin{equation}
{}^{(5)}G_{AB} + {}^{(5)}g_{AB}\, \Lambda_5 = 0 \,.
\end{equation}

The requirement of spatial homogeneity and isotropy on the brane is
sufficiently restrictive to allow one to solve for the bulk geometry
for an arbitrary brane cosmology~\cite{BDEL}. For a non-static
Friedmann-Robertson-Walker (FRW) cosmology on the brane the bulk
geometry must be either (anti-)de Sitter or Schwarzschild-(anti-)de
Sitter~\cite{MSM,BCG,Gergely}.  Thus the expansion of a FRW universe
may be re-interpreted in the brane-world scenario as due to motion in a
static bulk.
The four-dimensional matter fields determine the brane
trajectory in the bulk spacetime via the junction condition by
producing the jump in the extrinsic curvature at the brane. Without
loss of generality, the surface energy-momentum on the brane can be
split into two parts, $T_{\mu\nu}-\lambda g_{\mu\nu}$, where
$T_{\mu\nu}$ is taken to be the matter energy-momentum tensor and
$\lambda$ a constant brane tension. The junction condition is
then~\cite{israel,BDL,SMS}
\begin{equation}
\label{Israel}
 \left[K^\mu_\nu\right] \equiv K^{\mu~+}_\nu -
K^{\mu~-}_\nu =
 -\kappa_5^{~2} \left( T^\mu_\nu-{1\over 3} g^\mu_\nu \left( T -
     \lambda \right) \right),
\end{equation}
where the extrinsic curvature of the brane is denoted by 
\be
\label{defKmunu}
K_{\mu\nu}= g_\mu^{~C} g_\nu^{~D} \left({}^{(5)}\nabla_C n_D\right) \,,
\ee
where ${}^{(5)}\nabla_C$ is the 5D covariant derivative.
We will consider the case where the brane is located at a
Z$_2$-symmetric orbifold fixed point. Imposing the Z$_2$ symmetry
in Eq.~(\ref{Israel}), we obtain
\begin{equation}
\label{boundary}
K^{\mu~+}_\nu = - K^{\mu~-}_\nu =
-{\kappa_5^{~2}\over 2} \left( T^\mu_\nu-{1\over 3} g^\mu_\nu
\left( T - \lambda \right) \right).
\end{equation}
Henceforth we will only refer to quantities on the `$+$' side of the
brane and will drop the $+$ superscript.
%
The Codazzi equation for a vacuum bulk,
\begin{equation}
\label{Codazzi}
\nabla_\mu K^\mu_\nu - \nabla_\nu K = 0 \,,
\end{equation}
together with Eq.~(\ref{boundary}) enforces local energy-momentum
conservation on the brane,
\begin{equation}
\nabla_\mu T^\mu_\nu=0 \,,
\end{equation} 
where $\nabla_\mu$ is the 4D covariant derivative.


Shiromizu, Maeda and Sasaki \cite{SMS} showed that the effective
four-dimensional Einstein equations in a brane world can be
obtained by projecting the five-dimensional equations onto the
4-dimensional brane.
Using the Gauss and Codacci equations, we can obtain the Einstein
tensor for the four-dimensional metric induced on the brane
\begin{equation}
{}^{(4)}G_{\mu\nu}
 = - {1\over 2} \Lambda_5
+ KK_{\mu\nu} -K^{~\alpha}_{\mu}K_{\nu\alpha} -{1 \over 2}g_{\mu\nu}
  \left(K^2-K^{\alpha\beta}K_{\alpha\beta}\right) - E_{\mu\nu},
\label{4dEinstein}
\end{equation}
where
\begin{equation}
E_{\mu\nu} \equiv {}^{(5)}C^E_{~A F B}n_E n^F g_\mu^{~A}
g_\nu^{~B} \, . \label{Edef}
\end{equation}
is the projected five-dimensional Weyl tensor.

The effective four-dimensional Einstein equations can then be
written using Eqs.~(\ref{boundary}) and~(\ref{4dEinstein})
\begin{eqnarray}
\label{modEinstein}
{}^{(4)}G_{\mu\nu}+\Lambda_4 g_{\mu\nu}
 = \kappa_4^2 \T_{\mu\nu}+\kappa_5^4\,\Pi_{\mu\nu} -E_{\mu\nu}\,,
\label{eq:effective}
\end{eqnarray}
%
where we define
\begin{eqnarray}
\Lambda_4 &=&\frac{1}{2} \Lambda_5 +\frac{\kappa_5^4}{12}\,\lambda^2 \,,
\label{Lambda4}\\
\kappa_4^2 &=& {\kappa_5^4\,\over6}\lambda\,,
\label{GNdef}\\
\Pi_{\mu\nu}&=& -\frac{1}{4} \T_{\mu\alpha}\T_\nu^{~\alpha}
+\frac{1}{12}\T\T_{\mu\nu}
+\frac{1}{8}g_{\mu\nu}\T_{\alpha\beta}\T^{\alpha\beta}-\frac{1}{24}
g_{\mu\nu}\T^2\,. \label{pidef}
\end{eqnarray}
It is clear that we require $\lambda>0$ to have a positive induced
gravitational coupling constant ($\kappa_4^2>0$), and then
$\Lambda_5<0$ to have $\Lambda_4=0$.

The power of this approach is that the above form of the 4D
effective equations of motion is independent of the evolution of
the bulk spacetime, being given entirely in terms of quantities
defined on, or near, the brane. Thus these equations apply to
brane-world scenarios with infinite or finite bulk, static or
evolving.
A limitation is that this may leave terms which are not completely
determined by the local dynamics on the brane~\cite{SMS,Roy}. In
particular the tensor $E_{\mu\nu}$ is due to the 5-dimensional Weyl
tensor and so can be affected by gravitational waves propagating in
the bulk. In the case where the induced 4D metric is required to be
homogeneous and isotropic this strong symmetry constrains the
traceless tensor, $E_{\mu\nu}$, to be covariantly conserved on the
brane, so that it acts like non-interacting radiation. However we will be
interested in arbitrary linear perturbations about spacetimes
including an FRW brane and then we will need information from the full
five-dimensional field equations to describe the evolution of
inhomogeneous perturbations on the brane.  Nonetheless the dynamics
and effective gravity on the brane can be interpreted, and often most
easily understood, in terms of the effective four-dimensional Einstein
equations~(\ref{modEinstein}).

In section~\ref{Sbackground} we briefly summarise essential results
for the evolution of FRW brane-world cosmologies in a
Schwarzschild-anti-de Sitter bulk which we shall use as our
unperturbed background solution. In section~\ref{Sperturb} we
introduce linear perturbations of the bulk metric split into scalar,
vector and tensor perturbations (defined with respect to 3D spatial
hypersurfaces) in an arbitrary gauge. It is trivial to construct
quantities independent of the 3D-spatial gauge, and this is sufficient
to give gauge-invariant vector and tensor perturbations, but there are
alternative choices of temporal and bulk gauges and hence different
choices of gauge-invariant scalar perturbations.
In section~\ref{S5Dlong} we define gauge-invariant quantities
coinciding with metric perturbations in the 5D longitudinal gauge,
whereas in section~\ref{SGN} we construct quantities in a Gaussian
normal gauge where the 5D bulk metric perturbations coincide with the
induced 4D metric perturbations on the brane. This is particularly
suited for imposing the perturbed junction conditions, given in
section~\ref{Smatter}, which allows us to interpret the effect of bulk
metric perturbations as seen by the brane-bound observer in
section~\ref{Sview}. We show in section~\ref{SWeyl} that one can
construct gauge-invariant perturbations in order to describe the
evolution of the Weyl tensor perturbations throughout the bulk.
Finally we summarise our results in section~\ref{Ssummary}.
To assist the reader we have given the full unexpurgated tensor
components in appendices and included in the main text only simplified
expressions.

\section{Cosmological background solution}
\label{Sbackground}

In order to study inhomogeneous bulk metric perturbations we will
pick a specific form for the unperturbed 5D spacetime that
accommodates any spatially flat FRW cosmological solution on the
brane at $y=0$,
\begin{equation}
\label{backmetric}
ds_5^2 = - n^2(\eta,y) d\eta^2 + a^2(\eta,y)
\delta_{ij}dx^i dx^j
 + dy^2 \,.
\end{equation}
This form, using a Gaussian normal bulk coordinate, $y$, which
measures the proper distance from the brane, has been extensively
studied in the literature and includes anti-de Sitter
spacetime as a special case.
In particular it was shown in Ref.~\cite{BDEL} that it is possible to
use the 5D Einstein equations to
solve exactly for the dependence of the metric upon the bulk
coordinate $y$ in order to write (for ${}^{(5)}\Lambda<0$)~\cite{BDEL}
\begin{equation}
a^2 (\eta,y) = \alpha(\eta) \cosh(\mu |y|) + \beta(\eta) \sinh(\mu
|y|)
- \gamma^2(\eta)
 \,,
\end{equation}
where $\mu^2=-2\Lambda_5/3$ describes the curvature of the 5D space,
and $\alpha$, $\beta$ and $\gamma$ are functions only of time that are
determined by the matching conditions at the brane.
Equation~(\ref{G50}) requires $(\dot a/n)'=0$, and hence we can
write
\begin{equation}
n(\eta,y) = {\dot{a} \over a_b H} \,,
\end{equation}
where $H(\eta)=\dot{a}_b/n_ba_b$ is the Hubble expansion on the brane
and we use a dot to denote derivatives with respect to coordinate time
$\eta$, prime to denote derivatives with respect to the bulk
coordinate $y$, and the subscript $b$ to denote bulk quantities
evaluated at the brane.

However the evolution of the FRW cosmology on the brane can be
determined solely from the local 5D Einstein equations at the brane.
The 4D components of the 5D Einstein tensor [Eqs.~(\ref{G00})
and (\ref{Gij})] can be written in terms of the 4D Einstein tensor
[Eqs.~(\ref{4DG00}) and (\ref{4DGij})] as 
\bea 
\label{simpleG00}
{}^{(5)} G^0_0 &=& {}^{(4)}G^0_0 + E^0_0 + 3\frac{\ab^{\prime
2}}{\ab^2} 
 = - \Lambda_5 \,, \\ 
\label{simpleGij}
{}^{(5)}G^i_j &=& {}^{(4)}G^i_j + E^i_j +
\left( \frac{\ab^{\prime 2}}{\ab^2} + 2\frac{\ab' \nb'}{\ab
\nb}\right) \delta^i_j = -\Lambda_5 \delta^i_j \, ,
\eea
where 
the projected 5D Weyl tensor is given from Eqs~(\ref{E00}) and
(\ref{Eij}) by
\begin{eqnarray}
\label{backE00}
E^0_{~0} &=& 3\frac{a_b''}{a_b} + {1\over2} \Lambda_5\, , \\
E^i_{~j} &=& -\frac{1}{3} E^0_{~0} \delta^i_{~j} \, ,
\end{eqnarray}
and we have simplified these expressions using the 5D Einstein equations
(\ref{G55}),(\ref{G00}), and (\ref{Gij}) which give the constraint equation
\be
\label{backconstraint}
\frac{n''}{n} +3 \frac{a''}{a} = - {2\over3}\Lambda_5 \,.
\ee

The extrinsic curvature, defined in Eq.~(\ref{defKmunu}) above, 
of the FRW brane is
%
%
\bea
\label{K00}
K^0_0 &=& \frac{n'_b}{n_b} \,, \\
K^i_j &=& \frac{a'_b}{a_b} \,.
\eea
The energy-momentum tensor in a background with homogeneous density
$\rho$ and pressure $P$ is given by
\bea
T^0_0 &=& -\rho \,, \\
\label{Tij}
T^i_j &=& P \, \delta^i_j \,.
\eea
Hence the junction conditions given in Eq.~(\ref{boundary}), using
Eqs.~(\ref{K00}--\ref{Tij}) for a FRW brane, require
\bea
\label{n'/n}
\frac{n_b'}{n_b} &=&
\frac{\kappa^2_5}{2} \left(\frac{2}{3} \rho + P
-\frac{1}{3} \lambda \right) \,, \nonumber\\
\label{a'/a}
\frac{a_b'}{a_b} &=& -\frac{\kappa^2_5}{2} \left(\frac{1}{3} \rho
+ \frac{1}{3} \lambda\right) \,.
\eea

Thus the 5D Einstein equations (\ref{simpleG00}) and (\ref{simpleGij})
at the brane are of the general form given in Eq.~(\ref{modEinstein})
for the 4D effective Einstein equations where $\Lambda_4$ is given by
Eq.~(\ref{Lambda4}) and the quadratic correction to the effective
energy-momentum tensor of ordinary matter, $\Pi^{\mu}_{\nu}$ defined in
Eq.~(\ref{pidef}), is given by
\bea 
\Pi^0_0 &=& -\frac{1}{12} \rho^2 \,, \\ 
\Pi^i_j &=& \frac{1}{12}\rho(\rho+2P) \delta^i_j \,. 
\eea

Note that local energy conservation follows from the 5D Einstein equation
(\ref{G50}), using the junction conditions (\ref{a'/a})
\be
\label{backenergy} 
\dot\rho=-3\frac{\dot a}{a}(\rho+P) \,. 
\ee
One can verify that the quadratic energy-momentum tensor $\Pi^\mu_\nu$
is also conserved for a spatially homogeneous fluid~\cite{SMS}. The 4D
Bianchi identities, $\nabla_\mu {}^{(4)}G^\mu_\nu=0$ then ensure that the
contribution of the 5D Weyl tensor to the effective 4D Einstein
equations~(\ref{modEinstein}), $E^\mu_\nu$, is also conserved, and
thus the FRW brane dynamics are completely determined once one has set
the initial conditions for all these terms on the brane, without
requiring any further knowledge of the bulk behaviour away from the
brane.

\section{Linear perturbations}
\label{Sperturb}

We now proceed to consider arbitrary linear perturbations about the
background metric defined in Eq.~(\ref{backmetric}). In keeping with
the standard approach in cosmological perturbation
theory~\cite{Bardeen} we will introduce scalar, vector and tensor
perturbations defined in terms of their properties on the 3-spaces at
fixed $t$ and $y$ coordinates.

We can write the most general perturbed metric to first-order
as
\begin{equation}
\label{pertmetric} g_{AB}= \left(
\begin{array}{ccc}
-n^2(1+2\A) & a^2(B_{,i} - \S_i) & n\Ay \nonumber\\ a^2(B_{,j} -
\S_j) & a^2\left[ (1+2\R)\delta_{ij} + 2E_{,ij} + F_{i,j} +
F_{j,i} + h_{ij} \right] & a^2(B_{y,i} - \S_{yi})
 \nonumber\\
n\Ay & a^2(B_{y,i} - \S_{yi}) & 1+2\Ayy \nonumber
\end{array}
\right) \, ,
\end{equation}
where $\A$, $B$, $\R$, $E$, $\Ay$ and $\Ayy$ are scalars, $\S_i$,
$F_i$, and $\S_{yi}$ are (divergence-free) 3-vectors, and $h_{ij}$ is a
(transverse and traceless) 3-tensor. The reason for splitting the
metric perturbation into these three types is that they are
decoupled in the linear perturbation equations. 

In the perturbed spacetime there is a gauge-dependence in the
definition of the scalar and vector perturbations under a
first-order coordinate transformation, $x^A\to x^A+\xi^A$, which
we will write as
\begin{eqnarray}
\label{shift}
\eta &\to& \eta + \delta\eta \, ,\nonumber \\
x^i &\to& x^i + \delta x_,^{~i} + \delta x^i \,, \\
y &\to& y + \delta y \, , \nonumber
\end{eqnarray}
where $\delta\eta$, $\delta x$ and $\delta y$ are scalars and
$\delta x^i$ is a (divergence-free) 3-vector.
\subsubsection{Scalars}
The scalar perturbations transform as
\begin{eqnarray}
\label{transform}
\A &\to& \A - \dot{\delta\eta} - \frac{\dot
n}{n}\delta\eta -
\frac{n'}{n} \delta y \, , \nonumber \\
\R &\to& \R - \frac{\dot a}{a} \delta\eta
- \frac{a'}{a} \delta y \, , \nonumber \\
B &\to& B + \frac{n^2}{a^2} \delta \eta - \dot{\delta x} \, , \\
B_y &\to& B_y - \delta x' - \frac{1}{a^2} \delta y \, , \nonumber \\
E &\to& E - \delta x \, , \nonumber \\
\Ay &\to& \Ay +n\delta\eta' - \frac{1}{n} \dot{\delta y}\, , \nonumber \\
\Ayy &\to& \Ayy - \delta y' \, . \nonumber
\end{eqnarray}
%
%
To eliminate the spatial gauge dependence we introduce the
spatially gauge-invariant combinations
\be \label{scalarGI}
\begin{array}{lcl}  \sigma &\equiv& -B +
\dot{E} \, ,\\
\sigma_y &\equiv& -B_y + E' \, ,
\end{array}
\ee
which are subject only to temporal and bulk gauge transformations
\begin{eqnarray}
\label{sigmatransforms}
\sigma &\to& \sigma - {n^2\over a^2} \delta\eta \,, \nonumber \\
\sigma_y &\to& \sigma_y + {1\over a^2}\delta y \,.
\end{eqnarray}
The variable $\sigma$ represents the shear of a unit timelike vector
field projected onto the brane, whereas $\sigy$ represents the shear
with respect to a unit spacelike vector field orthogonal to the brane.
We sometimes also use the spatially-gauge invariant
combination
\be
\label{sgiB}
\B \equiv B'- \dot{B_y} = \dot{\sigma}_y - \sigma' \, . 
\ee

\subsubsection{Vectors}
The vector perturbations transform as
\begin{eqnarray}
\S_i &\to& \S_i + \dot{\delta x_i} \, ,  \\
\S_{yi} &\to& \S_{yi}
+ \delta {x_i}' \,, \nonumber \\
F_i &\to& F_i -\delta x_i \, .
\nonumber
\end{eqnarray}

It will be convenient to separate out time and bulk variation of vector
perturbations from their spatial dependence:
\begin{eqnarray}
\S_i &=& \S(t,y) \ei(x) \, ,  \\
\S_{yi} &=& \S(t,y) \ei(x) \,, \nonumber \\
F &=& F(t,y) \ei(x) \, ,
\nonumber
\end{eqnarray}
where the vector $\ei(x)$ is divergence-free ($\hat{e}_{i,}^{~i}=0$).
We can then introduce the gauge-invariant combinations
\be
\label{vectorGI}
\begin{array}{lcl}
\t &\equiv& \S + \dot{F} \, , \\
\ty &\equiv& \Sy + F' \, , \\
\s &\equiv& \S' - \dot{\Sy} = \t'-\dot\ty \, ,
\end{array}
\ee
which eliminates any gauge-ambiguity for the vector perturbations.

\subsubsection{Tensors}
The tensor perturbations $h_{ij}$ are automatically gauge-invariant.
\begin{eqnarray}
h_{ij} &\to& h_{ij} \, .
\end{eqnarray}


\section{Bulk perturbations in the 5D-longitudinal gauge}
\label{S5Dlong}

\setcounter{subsubsection}{0}

The bulk and temporal gauges are fully determined by setting
\bea
\label{def5Dlong}
\tilde\sigma &=& 0 \, , \nonumber \\
\tilde\sigy &=& 0 \,,
\eea
which has been termed the longitudinal gauge~\cite{cvdb}. We will refer
to this as the 5D-longitudinal gauge, to avoid later confusion with
quantities in the 4D-longitudinal gauge on the brane.

{}From an arbitrary gauge, we can obtain quantities in the
5D-longitudinal gauge, using Eqs.~(\ref{sigmatransforms})
and~(\ref{def5Dlong}), to define the specific gauge transformation
\bea
\label{5Dlongshift}
\delta\tilde\eta = {a^2\over n^2} \sigma \,,\nonumber\\
\delta\tilde{y} = - a^2 \sigma_y \,.
\eea
Hence, from Eqs.~(\ref{transform}) and (\ref{5Dlongshift}),
we can define the remaining metric perturbations in the
5D-longitudinal gauge as
\bea
\label{5Dlongpert}
\lA &=& A- \frac{1}{n} \left(\frac{a^2}{n^2}\sigma\right)^\cdot
 + \frac{n'}{n} \left( a^2\sigy \right) \, , \nonumber \\
\lR &=& \R -\frac{\dot a}{a} \frac{a^2}{n^2} \sigma +\frac{a'}{a}a^2
\sigy \, , \nonumber \\
\lAy &=& \Ay +n\left(\frac{a^2}{n^2}\sigma\right)'
+\frac{1}{n}\left(a^2\sigy\right)^\cdot \, , \nonumber \\
\lAyy &=& \Ayy +\left(a^2\sigy\right)' \, .
\eea
These are equivalent to the gauge-invariant bulk perturbations
originally introduced in covariant form by
Mukohyama~\cite{Mukoh,Kodama} and in a coordinate-based approach by
van den Bruck et al~\cite{cvdb}.

\subsubsection{Scalar perturbations}

First note that the spatial trace part of the 5D Einstein equations,
given in Eq.~(\ref{5dtrace}), simplifies in the 5D-longitudinal gauge
to
\be
\lA + \lR + \lAyy=0 \,.
\ee
We can therefore eliminate any one of these metric perturbations
from the equations. This simple constraint has also been found to hold
for the case of static bulk solutions with a scalar field in the bulk
\cite{Veneziano} and continues to hold in a cosmological setting in
the absence of anisotropic stress in the bulk \cite{cvdb}. 

Mukohyama \cite{Mukoh} (see also Ref.~\cite{Kodama}) was the first to
show that the perturbed 5D Einstein equations (\ref{5D0i}),
(\ref{5Dij}) and (\ref{5D4i}) in the absence of bulk matter
perturbations ($^{(5)}\delta G^A_B=0$) are solved in an anti-de Sitter
background if the metric perturbations are derived from a ``master
variable'', $\Omega$:
\bea
\label{lA}
\lA &=& 
-\frac{1}{6a}\left\{2\Omega''-\frac{n'}{n}\Omega'
+\frac{\Lambda}{6}\Omega
+\frac{1}{n^2}\left(\ddot\Omega-\frac{\dot n}{n}\dot\Omega\right)\right\}
\, , \\
\label{lAy}
\lAy &=& 
\frac{1}{na}\left(\dot\Omega'-\frac{n'}{n}\dot\Omega\right)\,, \\
\label{lAyy}
\lAyy &=& 
\frac{1}{6a}\left\{
\Omega''-2\frac{n'}{n}\Omega'
+\frac{2}{n^2}\left(\ddot\Omega-\frac{\dot n}{n}\dot\Omega\right)
-\frac{\Lambda}{6}\Omega \right\} \,, \\
\label{lR}
\lR &=& 
\frac{1}{6a}\left\{
\Omega''+\frac{n'}{n}\Omega'
-\frac{1}{n^2}\left(\ddot\Omega-\frac{\dot n}{n}\dot\Omega\right)
+\frac{\Lambda}{3}\Omega\right\} \,.
\eea
{}The remaining perturbed 5D Einstein Eqs.~(\ref{5D00}), (\ref{5D04})
and~(\ref{5D44}) then yield a single wave equation governing the
evolution of the master variable $\Omega$ in the bulk:
\be
\label{scalarmastereom}
\left( {1\over na^3} \dot\Omega \right)^\cdot
+ \left( {\Lambda_5 \over6} - {\nabla^2\over a^2} \right) {n\over a^3}
\Omega
=
\left( {n\over a^3} \Omega' \right)^\prime
 \,.
\ee

This 5D wave equation is separable only for a separable background
solution which corresponds to the case of a de Sitter
brane-world~\cite{GS,BMW}.

\subsubsection{Vector perturbations}

As for scalar perturbations, the evolution equation for vector metric
perturbations, 
Eq.~(\ref{5dvecGij}), is automatically satisfied when the gauge
invariant vector perturbations $\tau$ and $\tau_y$ are derived from a
single ``master variable"~\cite{Mukoh,Kodama,BMW}
\bea \tau &=& \frac{n}{a^3} \Omega^{\prime (vector)}, \\ \tau_y
&=& \frac{1}{na^3} \dot{\Omega}^{(vector)}. \eea
The remaining perturbed 5D Einstein equations~(\ref{5dvecG0i}) and
~(\ref{5dvecG4i}) then yield a single governing equation
for $\Omega^{(vector)}$,
\be 
\label{vectorOmegaeom}
\frac{1}{n^2} \left[ \ddot{\Omega}^{(vector)} -
\left(3\frac{\dot{a}}{a} + \frac{\dot{n}}{n} \right)
\dot{\Omega}^{(vector)} \right] - \left[ \Omega^{\prime \prime
(vector)} - \left(3\frac{a'}{a} - \frac{n'}{n} \right)
\Omega^{\prime (vector)} \right] + \frac{k^2}{a^2}
\Omega^{(vector)} = 0 \,, 
\ee
for each Fourier mode, $\nabla^2\ei=-k^2\ei$.

\subsubsection{Tensor perturbations}

The only Einstein equation for the gauge-invariant tensor
perturbations, Eq.~(\ref{5Dtensor}), already has the form of the wave
equation for a master variable. This can be made explicit if we
separate out the 3-space dependence, writing
\be
h_{ij}(t,x^i,y) = \Omega^{(tensor)}(t,y) \hat{e}_{ij}(x^i) \,,
\ee
where $e_{ij}(x^i)$ is a transverse, tracefree harmonic on the
spatially flat 3-space, $\nabla^2\hat{e}_{ij}=-k^2\hat{e}_{ij}$. 
Equation~(\ref{5Dtensor}) then yields
\be
\label{heom}
\frac{1}{n^2} \left[ \ddot{\Omega}^{(tensor)} +
\left(3\frac{\dot{a}}{a} -
\frac{\dot{n}}{n}\right)\dot{\Omega}^{(tensor)}\right] - \left[
\Omega^{\prime \prime (tensor)} + \left(3\frac{a'}{a} +
\frac{n'}{n}\right) \Omega^{\prime (tensor)} \right] +
\frac{k^2}{a^2}\Omega^{(tensor)} = 0 \,.
\ee
This equation and its solutions for the special case of a de Sitter
brane were studied in Ref.~\cite{LMW}.

\section{The Gaussian normal gauge}
\label{SGN}

An alternative choice of gauge often used \cite{GS,Langlois,Sasaki} is
a Gaussian normal (GN) gauge, where the bulk $y$ coordinate measures
the proper distance from the brane, up to and including first-order
perturbations. This requires the metric perturbations $B_y$, $\Ay$ and
$\Ayy$ to vanish. Starting from an arbitrary 5D gauge, this can be
enforced throughout the bulk via a coordinate shift
$x^A\to\bar{x}^A=x^A+\bar{\delta x}^A$, [see Eqs.~(\ref{shift})
and~(\ref{transform})] such that,
\begin{eqnarray}
\label{GNshift}
\bar{\delta\eta}' &=&  {1\over n^2} \dot{\delta\bar{y}}
 - {1\over n} \Ay \,,\nonumber \\
\bar{\delta y}' &=& \Ayy \,, \nonumber\\
\bar{\delta x}' &=& - {1\over a^2} \delta\bar{y} + B_{y} \,, \\
\bar{\delta x}_i' &=& - \S_{yi} \nonumber \,.
\end{eqnarray}
This does not completely fix the gauge, but only the
$y$-derivative of the gauge-transformation, 
analogous to the synchronous gauge in conventional 4D perturbation
theory.

The projected metric on any constant-$y$ hypersurface is then
\footnote{In principle the choice of coordinates, and hence
  gauge transformations, on the brane may be chosen quite
  independently of the bulk~\cite{Mukoh,Deruelle}.
  However in order to relate the bulk perturbations as directly as
  possible to the brane-world observer we will use the same
  coordinates $(t,x^i)$ on the brane as are used to follow
  perturbations in the bulk.}
\begin{equation}
\label{GNpertmetric}
g_{AB}= {}^{(5)}g_{AB} - n_A n_B
 = \left(
\begin{array}{ccc}
-n^2(1+2\bar{\A}) & a^2(\bar{B}_{,i} - \bar{\S}_i) & 0 \nonumber\\
a^2(\bar{B}_{,j} - \bar{\S}_j) & a^2\left[
(1+2\bar{\R})\delta_{ij} + 2\bar{E}_{,ij} + \bar{F}_{i,j} +
\bar{F}_{j,i} + \bar{h}_{ij} \right] & 0 \nonumber\\ 0 & 0 & 0
\nonumber
\end{array}
\right) \,,
\end{equation}
where we denote by a bar the metric perturbations in a Gaussian
normal gauge.

\subsection{Transverse-tracefree GN gauge}

A technique commonly used to study the propagation of gravitational
waves in a vacuum spacetime is to work in a gauge in which the
perturbations are transverse and tracefree in the background
spacetime. We shall now show that while the transverse and
tracefree condition is certainly sufficient to fix the residual
gauge-freedom in a Gaussian normal gauge, it actually over constrains
the problem for a general cosmology and its usefulness is restricted
to the cases of a maximally symmetric 4D (anti-)de Sitter brane.

The linearly perturbed five-dimensional Ricci tensor can be 
written as
\be 
^{(5)}\delta R_{BD}=
\frac{1}{2}\left\{^{(5)}\nabla^A\left(^{(5)}\nabla_D \,^{(5)}\delta g_{BA}
    +^{(5)}\nabla_B \,^{(5)}\delta g_{DA}\right)-^{(5)}\nabla^A\,
  ^{(5)}\nabla_A \,^{(5)}\delta g_{BD} -^{(5)}\nabla_D
  \,^{(5)}\nabla_B \,^{(5)}\delta g^A_A \right\} \,, 
\ee
where $^{(5)}\nabla_A$ is the five-dimensional covariant derivative.
Swapping the order of differentiation of the first two terms, this can
be rewritten as 
\bea
^{(5)}\delta R_{AB}
&=& \frac{1}{2}\left( ^{(5)}\nabla_A\,^{(5)}\nabla^C
  \,^{(5)}\delta g_{BC} + 
^{(5)}\nabla_B\,^{(5)}\nabla^C \delta g_{AC} -
^{(5)}\nabla_A\,^{(5)}\nabla_B \,^{(5)}\delta g^C_C  
-^{(5)}\nabla_C\,^{(5)}\nabla^C \,^{(5)}\delta g_{AB} \right) 
\nonumber\\
&& \quad +^{(5)}R_{CADB} \,^{(5)} \delta g^{CD} \,.
\eea
In the absence of any bulk matter perturbations,
the perturbed 5D Einstein equations require $^{(5)}\delta R_{AB} =0$.
Thus when the 5D perturbations are transverse
($^{(5)}\nabla^C\,^{(5)}\delta g_{AC}=0$) and tracefree ($^{(5)}\delta
g^C_C=0$),  the perturbed Einstein equations can be written as
\begin{equation}
^{(5)}\Box ^{(5)}\delta g_{AB}
 = 2 \, ^{(5)}R_{CADB} \,^{(5)}\delta g^{CD} \,,
\end{equation}
where $^{(5)}\Box=^{(5)}\nabla_C\,^{(5)}\nabla^C$. 
The background Riemann tensor in AdS$_5$ is given by
\be
^{(5)}R_{ABCD}=\frac{\Lambda_5}{6} \left(
  ^{(5)}g_{AC}\,^{(5)}g_{BD}-^{(5)}g_{AD}\,^{(5)}g_{BC} \right) \,.
\ee
%
%
Therefore to linear order in the metric perturbations and enforcing
the transverse and traceless conditions the field equations in the
absence of matter are given by
\be
\label{boxgAB}
^{(5)}\Box ^{(5)}\delta g_{AB} =- \frac{1}{3} \Lambda_5 \,^{(5)}\delta
g_{AB} \,. 
\ee

\subsubsection{Scalar perturbations}

The tracefree condition, in a GN gauge, requires
\be
\label{tracefreeGN}
A+3\R+\nabla^2 E=0 \,.
\ee
The transverse condition in general gives rise to four constraint
equations, which can be written, using Eq.~(\ref{tracefreeGN}) above, as
\bea
\label{transverseGN}
&& \nabla^2 B +2\left(\dot A + 4\frac{\dot a}{a}A \right) =0 \,, \\
%
&& \left\{
\frac{a^2}{n^2}\left[\left(\frac{\dot n}{n}-5\frac{\dot a}{a}\right)B-
\dot B\right] -2A -4\R \right\}_{,i} = 0 \,, \\
%
&& 2\left(\frac{a'}{a} - \frac{n'}{n} \right) A = 0\,.  
\eea 
Unless $(a/n)'=0$, i.e., $a$ and $n$ have the same $y$-dependence, the
five constraint equations require that the four Gaussian normal scalar
metric perturbations are all identically zero. Using the background
field equation $^{(5)}G^4_0=0$ [see Eq.~(\ref{G50})], this
implies that it is only possible to use the transverse and
tracefree Gaussian normal gauge for a separable bulk metric, which
corresponds to an (anti-)de Sitter or Minkowski brane.

Quite generally, one can see that the gauge shift required to impose
the traceless (\ref{tracefreeGN}) and transverse (\ref{transverseGN})
conditions throughout the bulk, will only be compatible with the
Gaussian normal gauge shift (\ref{GNshift}) if the metric
perturbations are composed of separable functions of brane coordinates
$\eta$ and $x^i$ and bulk coordinate $y$. And this requires that the
background bulk metric is separable.

Thus, only in the special case of a (anti-)de Sitter or Minkowski
brane, the wave equation~(\ref{boxgAB}) gives an evolution equation
for the scalar metric perturbation:
\be
\label{Aeom}
\frac{1}{n^2}\left\{\ddot A 
-\left(\frac{\dot n}{n}-7 \frac{\dot a}{a} \right)\dot A \right\}
+ \left[ 8\left(\frac{\dot a}{an}\right)^2   
+2\left(\frac{a'}{a}\right)^2 - \frac{1}{6} \Lambda_5 \right] A 
-\frac{1}{a^2}\nabla^2 A
=  A''+  4\frac{a'}{a} A \,,
\ee
and the remaining scalar metric perturbations can be deduced from the
constraint Eqs.~(\ref{tracefreeGN}) and (\ref{transverseGN}).

\subsubsection{Vector perturbations}

The transverse condition for the vector perturbations gives
a single constraint equation,
\be
\nabla^2 F_i+\frac{a^2}{n^2}\left[\dot S_i -\left(
\frac{\dot n}{n}-5\frac{\dot a}{a}\right) S_i \right]
=0 \,.
\ee
For this to hold throughout the bulk in a GN gauge,
Eq.~(\ref{GNshift}), restricts us to the special case of a separable
bulk/de Sitter brane.  The wave equation (\ref{boxgAB}) then yields a
decoupled equation of motion for the gauge-invariant vector
perturbation~\cite{BMW}
\be
{1 \over n^2} \left[ \ddot{\tau} + 3\left({\dot{a}\over a} -
    {\dot{n}\over n} \right) \dot\tau \right] + {k^2\over a^2} \tau =
\tau'' + 4{a'\over a} \tau'  \,,
\ee
which is consistent with the more generally applicable wave equation
for Mukohyama's master variable, $\Omega^{({\rm vector})}$, given in
Eq.~(\ref{vectorOmegaeom}). 

\subsubsection{Tensor perturbations}

Tensor perturbations, $h_{ij}$, that are transverse and tracefree on
3D spatial hypersurfaces, remain transverse and tracefree with respect
to the 4D metric. These perturbations are gauge-invariant and their
evolution equation is given in Eq.~(\ref{heom}).

\subsection{Brane-GN gauge}

As well as allowing arbitrary first-order metric perturbations in the
bulk, we should in general allow the coordinate location of the brane
to acquire a first-order perturbation, so that the brane is located at
$y_b=\xi(x^\mu)$.
The coordinate value of the brane location is a coordinate-dependent
quantity, transforming as
\be
\label{xitransform}
\xi \to \xi +\delta y_b \, ,
\ee
under the gauge-transformation given in Eq.~(\ref{shift}). It can
obviously be set to zero by a suitable choice of gauge.  Nonetheless
it is important to realise that such a gauge restriction is being
imposed when the coordinate location of the brane is fixed. It is not
possible to set the brane location to zero if the gauge has already
been completely fixed, e.g., in the 5D-longitudinal gauge or the
transverse-tracefree GN gauge, without
placing physical restrictions on the perturbations that can be
accommodated~\cite{cvdb,ivo}.
$\xi$ represents a 4D degree of freedom that although not part of the
intrinsic 5D metric perturbations, does appear in the 4D junction
conditions.
\begin{figure}
\begin{center}
\includegraphics[width=120mm]{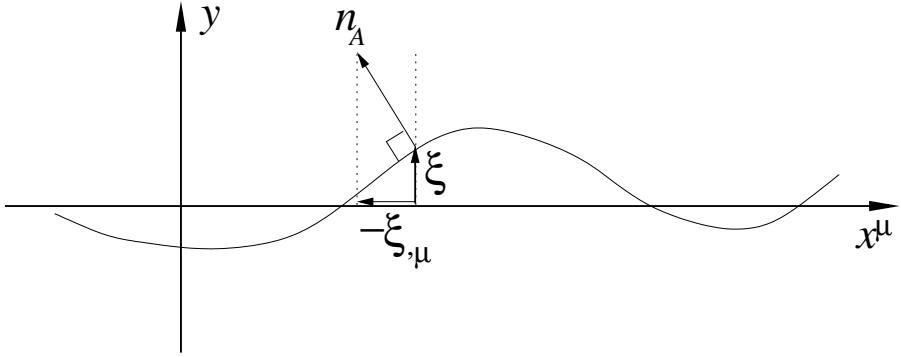} \\
\caption[branefig]{\label{branefig} The coordinate position $\xi$ of
the perturbed brane and the normal vector off the brane $n_A$.}
\end{center}
\end{figure}

The unit vector field orthogonal to the brane
then becomes
\be n_A=(-\xi_{,\mu},1+\Ayy) \,. \ee

Note that we will identify the brane location with the Z$_2$-symmetric
orbifold fixed point, and hence the covariant form of the junction
condition given in Eq.~(\ref{boundary}) to be imposed at the brane is
unchanged.
In practice, the junction conditions and Z$_2$ symmetry at the brane
are most easily presented in terms of the metric perturbations at the
brane in the Gaussian normal gauge, where the bulk coordinate $y$
still measures the proper distance from the brane including
first-order perturbations. 
%

To fix the coordinate position of the brane as $\bar{y}_b=\bar\xi=0$
requires in addition
\begin{equation}
\label{yshift}
\bar{\delta y}_b = -\xi \,.
\end{equation}
This then fixes completely $\delta y$, i.e., the bulk-slicing, but
leaves a residual temporal gauge-freedom, i.e., time-slicing, which
can be eliminated by fixing the time-slicing on any constant-$y$
hypersurface, e.g., $\bar{\delta\eta}_b$. This corresponds to the
standard 4D temporal gauge-dependence of the metric perturbations on
the brane.

The induced scalar metric perturbations on the brane are thus obtained
from arbitrary gauge perturbations via the transformations in
Eqs.~(\ref{shift}) using the specific coordinate shifts in
Eqs.~(\ref{GNshift}) and (\ref{yshift}), to give
\begin{eqnarray}
\label{GNpert}
\bar{\A}_b &=& \A_b + \frac{n_b'}{n_b} \xi \, , \nonumber \\
\bar{\R}_b &=& \R_b + \frac{a_b'}{a_b} \xi \, , \\
\bar{\sigma}_b &=& \sigma_b
 \, . \nonumber
\end{eqnarray}
These are the standard gauge-dependent metric perturbations
corresponding to lapse, curvature and shear perturbations in
conventional 4D cosmological perturbation theory~\cite{MFB}.
They transform in the standard way under temporal
gauge-transformations on the brane, $\eta_b\to\eta_b+\delta\eta_b$, to
give~\cite{Malik}
\begin{eqnarray}
\bar{\A}_b &\to& \bar{\A}_b - {1\over n_b} \left( n_b\delta\eta_b
\right)^\cdot \, , \nonumber \\
\bar{\R}_b &\to& \bar{\R}_b - {\dot{a}_b\over a_b} \delta\eta_b\, , \\
\bar{\sigma}_b &\to& \bar{\sigma}_b - {n_b^2\over a_b^2}\delta\eta_b
\, . \nonumber
\end{eqnarray}

One can recover the 4D-longitudinal gauge perturbations on
the brane by choosing $\delta\eta_b=(a_b/n_b)^2\bar{\sigma}_b$ in the
above equations, yielding $\bar{\sigma}_b\to0$ and
\begin{eqnarray}
\label{4Dlongpert}
\bar{\A}_b &\to& \Phi =  \bar{\A}_b - {1\over n_b} \left( {a_b^2\over
    n_b} \bar\sigma_b \right)^\cdot  \, , \nonumber \\
\bar{\R}_b &\to& -\Psi = \bar{\R}_b - {\dot{a}_b\over a_b} {a_b^2\over
    n_b^2} \bar\sigma_b \, ,
\end{eqnarray}
where $\Phi$ and $\Psi$ correspond to the 4D metric potentials defined
in Ref.~\cite{MFB} and are completely gauge-invariant. These
4D-longitudinal gauge perturbations on the brane do {\em not} in
general coincide with the 5D-longitudinal gauge perturbations.
Using the 5D-longitudinal gauge perturbations
$\tilde{\A}$ and $\tilde{\R}$ defined in Eq.~(\ref{5Dlongpert}) and
evaluated at the brane we obtain from Eqs.~(\ref{GNpert})
and~(\ref{4Dlongpert})
\begin{eqnarray}
\Phi =  \tilde{\A}_b + {n_b'\over n_b}\tilde\xi
  \, , \nonumber \\
\Psi = - \tilde{\R}_b - {a_b'\over a_b}\tilde\xi
\, ,
\end{eqnarray}
which will only coincide with the 5D-longitudinal gauge metric
perturbations if the brane location in the 5D-longitudinal
gauge remains unperturbed, i.e., $\tilde\xi=0$.

The remaining scalar bulk metric perturbation at the brane in a
Gaussian normal gauge is described by
\be
\bar{\sigma}_{yb} = \sigma_{yb} - \frac{1}{a^2} \xi \, ,
\ee
which is completely gauge-invariant,
but is not part of the induced metric perturbations on the brane.
Instead we shall show in Section \ref{Einsteinonthebrane}
that it appears as a source term in the
effective Einstein equations~(\ref{modEinstein}).
This can immediately be identified with the brane location in the
5D-longitudinal gauge, $\tilde\xi=-a^2\bar\sigma_{yb}$.
Note that in terms of the Gaussian normal metric perturbations
$\bar{E}$ and $\bar{B}$ in Eq.~(\ref{GNpertmetric}) we can write
$\bar\sigma_{y}=\bar{E}'$, while the alternative gauge-invariant
combination, defined in Eq.~(\ref{sgiB}), is given in Gaussian normal
gauge by
%
$\bar{\B} = \dot{\bar\sigma}_{y}-\bar\sigma'=\bar{B}'$.

The vector perturbations, $\t$ and $\ty$, and the tensor perturbations,
$h_{ij}$, are invariant under the transformations given in
Eqs.~(\ref{GNshift}) and (\ref{yshift}). $\t$ and $h_{ij}$ are the
induced 4D vector and tensor metric perturbations, but $\ty$, like
$\bar{\sigma}_y$, acts like a source term in the effective Einstein
equations~(\ref{modEinstein}).

%

In the following sections we shall show the central role played by
brane-GN gauge perturbations in interpreting the effect of bulk
perturbations on the brane. To minimise the notation we henceforth
drop the `$b$' subscripts for metric perturbations evaluated at the
brane.

\section{Matter perturbations}
\label{Smatter}

The energy-momentum tensor for arbitrary matter
perturbations on the brane is given by
\begin{eqnarray}
\label{dT00}
\delta\T^0_0 &=& -\delta\rho \,,\\
\label{dT0i}
\delta\T^0_i &=& \delta p_i  \,,\\
\label{dTij}
\delta\T^i_j &=& \delta P \, \delta^i_j + \delta\pi^i_j \,,\\
\label{dTtrace}
\delta\T &=& 3\delta P - \delta\rho \, ,
\end{eqnarray}
where $\delta p_i$ is the perturbed three-momentum,
$\delta\pi^i_j$ is the traceless anisotropic stress, and $\delta
T\equiv\delta T^{\mu}_{\mu}$ is the perturbed four-trace of the
energy-momentum tensor. The perturbed three-momentum can be decomposed
into scalar and vector components as
\be
\delta p_i \equiv \delta p_{,i} + \dpvec\,\ei \,,
\ee
and the anisotropic stress can be decomposed as
\be
\label{deltapidecomposed}
\delta\pi^i_j \equiv \left( \nabla^i\nabla_j
-\frac{1}{3}\delta^{i}_{j}\nabla^2\right) \delta \pi
+ \dpivec \eij + \dpitens \, ,
\ee
where  $\delta \pi$, $\dpivec$ and  $\dpitens$ are the scalar, vector
and tensor parts of the anisotropic stress tensor, respectively.

The matter quantities transform under the temporal gauge transformation,
$\eta\to\eta+\delta\eta$ on the brane, as~\cite{KS}
\bea
\delta\rho &\to& \delta\rho -\dot\rho\delta\eta \,, \nonumber\\
\delta P &\to& \delta P - \dot P \delta\eta \,, \\
\delta p &\to& \delta p + (\rho+P)\delta\eta \,,
\nonumber 
\eea
respectively, and the anisotropic stress tensor, $\delta\pi^i_j$,
and the vector momentum, $\dpvec\ei$, are gauge-invariant.

Equation~(\ref{boundary}) yields the orbifold boundary condition
for metric perturbations at the brane:
\begin{equation}
\label{boundarypert}
\delta K^{\mu}_\nu = -{\kappa_5^{~2}\over 2}
\left( \delta \T^\mu_\nu - {1\over 3} \delta^\mu_\nu \delta\T\right) \, ,
\end{equation}
which we again decompose into scalar, vector and tensor parts.

\subsubsection{Scalar perturbations}


The contribution of the scalar metric perturbations in the Gaussian
normal gauge to the extrinsic curvature of the brane is simply
given in terms of the normal derivatives:
\bea
\label{dK00}
\delta K^0_{~0} &=& \bar{\A} ' \, , \\
\delta K^0_{~i} &=& -\frac{\ab^2}{2\nb^2} \bar{\B}_{,i} \, , \\
\label{dKij}
\delta K^i_{~j} &=& \bar{\sigy}^{i}_{,~j} +
\delta^i_{~j} \bar{\R}'  \, . \eea

Substituting the gauge transformations (\ref{transform}) and
(\ref{sigmatransforms}) of the metric perturbations, using
Eqs.~(\ref{GNshift}) and~(\ref{yshift}), into the above expressions
(\ref{dK00}-\ref{dKij}) yields an expression for the extrinsic
curvature perturbations in an arbitrary gauge
\bea
\label{scalarkmunu2}
\delta K^0_{~0} &=& A'
+\frac{1}{\nb^2}\left[\ddot\xi- \frac{\dot \nb}{\nb}\dot\xi
+\nb\dot {\Ay}  \right] -\frac{\nb'}{\nb} A_{yy}
+\left(\frac{\nb'}{\nb}\right)'\xi \, , \\
\delta K^0_{~i} &=& -\frac{1}{2\nb^2}\left\{\ab^2 \B
-\nb\Ay-2\left(\dot\xi-\frac{\dot\ab}{\ab}\xi\right)
\right\}_{,i}
\, , \\
\delta K^i_{~j} &=&
\left\{\R'
+\frac{1}{\nb^2}\frac{\dot\ab}{\ab}\left(\dot\xi+n\Ay\right)
-\frac{\ab'}{\ab}\Ayy+\left(\frac{\ab'}{\ab}\right)'\xi
\right\} \,\delta^i_{~j}
+\left\{\sigma_y-\frac{1}{\ab^2}\xi\right\}^{~i}_{~,j} \, ,
\eea
which introduces additional terms involving time derivatives of the
metric perturbations and $\xi$.

The perturbed junction conditions in
Eq.~(\ref{boundarypert}) relate the extrinsic curvature
perturbations (\ref{dK00}--\ref{dKij}) to the matter
perturbations on the brane given in
Eqs.~(\ref{dT00}-\ref{dTtrace}). In the brane-GN gauge we find
\bea
\label{juncA}
\bar{\A}' &=& \frac{1}{6}\,\kappa^2_5\left(
2\delta\rho+3\delta P \right) \, , \\
\label{juncB}
\bar{\B} &=& \kappa^2_5\, \frac{\nb^2}{\ab^2}\delta p  \, , \\
\label{juncR}
\nabla^2\bar\sigy +3\bar\R' &=& -\frac{1}{2}\kappa^2_5\,\delta \rho \, , \\
\label{juncS}
\bar\sigy &=& -\frac{1}{2}\kappa^2_5\,\delta\pi \, ,
\eea
giving a direct relation between the metric perturbations and the
matter perturbations. But in any other gauge, these relations
involve time-derivatives of the perturbations and so are non-local in
time \cite{Kodama,Mukoh}. Only the anisotropic stress $\delta\pi$
[Eq.~(\ref{juncS})] yields a local boundary condition for $\xi$ and
$\sigma_y$ in an arbitrary gauge.

Two of the junction conditions can be interpreted as definitions of
brane density and momentum perturbations in terms of bulk metric
perturbations at the brane, whose evolution can be determined through
the 5D Einstein equations. However any non-adiabatic or anisotropic
pressure perturbation is not determined from the Einstein equations
and hence represent additional constraints on the evolution of the
bulk metric perturbations~\cite{Kodama}.

\subsubsection{Vector perturbations}

The non-zero contribution of the vector perturbations to the
perturbed extrinsic curvature is given by
%
%
%
%
%
\bea
\label{vecK0i}
\delta K^0_{~i} &=& \frac{\ab^2}{2\nb^2} \; \s \; \ei \, , \\
\delta K^i_{~j} &=& \frac{1}{2} \; \ty \; \eij  \, .
\eea
Equating these with the vector perturbations in the matter on the
brane, Eqs.~(\ref{dT0i}-\ref{dTij}), we find
\bea
\label{matvecK1}
\s &=& -\kappa_5^{~2}
\, \frac{\nb^2}{\ab^2}\dpvec \, , \\
\label{matvecK2}
\ty &=& -\kappa_5^{~2} \, \dpivec \, .
\eea
%
As the vector perturbations are gauge-invariant, the local form of the
boundary conditions remains unchanged in arbitrary bulk or temporal
gauges.

\subsubsection{Tensor perturbations}
%
The contribution of the tensor perturbations to the extrinsic
curvature tensor is
\be
\delta K^i_j = \frac{1}{2} h^{i~\prime}_{j}
 = -{\kappa_5^{~2}\over 2} \,\dpitens \,.
\ee

If, for example, matter perturbations on the brane are described by 
linear perturbations of a scalar field then $\delta K^{\mu}_{\nu}$ 
for the vector and tensor perturbations on the brane must vanish.

\section{The view from the brane}
\label{Sview}

\subsection{Einstein equations on the brane}
\label{Einsteinonthebrane}

In order to make predictions for the behaviour of perturbations
seen by the brane-world observer it is necessary to relate the
five-dimensional metric perturbations to perturbations in the
effective four-dimensional Einstein equations on the brane.
{}From Eq.~(\ref{modEinstein}) we see that perturbations of the
bulk gravitational field only enter via the projected
Weyl tensor, $E^\mu_\nu$, which contributes like an effective
energy-momentum tensor in the effective Einstein equations on the
brane.
We will define an effective Weyl-fluid energy density in the
background by 
\be
\label{Weyldensity}
\kappa^2_4 \widetilde{\rho} = E^0_0 = 3{\ab''\over \ab} +
{1\over2}\Lambda_5 \, ,
\ee
with isotropic pressure $E^i_j=-(1/3)E^0_0\delta^i_j$.

The perturbed Weyl-fluid yields effective density, pressure and momentum
perturbations, which by analogy with
Eqs.~(\ref{dT00}--\ref{dTtrace}), we write as
\bea
\label{Weylfluid}
\delta E^0_0 &=& \kappa^2_4\widetilde{\delta\rho} \,, \nonumber \\
\delta E^0_i &=& - \kappa^2_4 \widetilde{\delta p_i} 
 \,, \\
\delta E^i_j &=& - \kappa^2_4 \left( \widetilde{\delta P}\, \delta^i_j +
  \widetilde{\delta\pi^i_j} \right) \,, \nonumber
\eea
where the isotropic pressure perturbation is 
$\widetilde{\delta P}=\widetilde{\delta\rho}/3$
and we can decompose the Weyl momentum and the Weyl anisotropic stress
into scalar, vector and tensor parts according to
\bea
\widetilde{\delta p_i} &\equiv& \widetilde{\delta p}_{,i} +
\widetilde{\delta p}^{\rm{(vector)}} \,\ei \,, \\
\widetilde{\delta\pi^i_j} &\equiv& \left( \nabla^i\nabla_j
-\frac{1}{3}\delta^{i}_{j}\nabla^2\right) \widetilde{\delta \pi}
+\eij \, \widetilde{\delta\pi}^{\rm{(vector)}}
+ \widetilde{\delta\pi^i_j}^{\rm{(tensor)}} \,.
\eea
%


%
\subsubsection{Scalar perturbations}

The scalar contribution to the projected Weyl fluid perturbations
defined in Eqs.~(\ref{Weylfluid}) can be written
using the 5D Einstein equations, (\ref{5D00})-(\ref{5dtrace}), 
to simplify the perturbed projected Weyl tensor
given in equations, (\ref{scalE00})-(\ref{scalET}). In terms of
normal derivatives of the Gaussian normal metric
perturbations the Weyl fluid perturbations are~\cite{Langlois}:  
\bea
\label{dE00}
\kappa^2_4 \widetilde{\delta\rho} 
= 3 \kappa^2_4 \widetilde{\delta P}
&=& -\left(
\bar{\A}''+2\frac{\nb'}{\nb}\bar{\A}' \right) \,, \\
\label{dE0i}
\kappa^2_4 \widetilde{\delta p} 
&=& - {1\over2} \frac{\ab^2}{\nb^2}\left\{ \bar{\B}' +
  \left(3\frac{\ab'}{\ab}-\frac{\nb'}{\nb}\right)\bar{\B}\right\} 
\,, \\
%
\label{dET}
\kappa^2_4 \widetilde{\delta \pi}
&=& \left( \bar{\sigy}'+2\frac{\ab'}{\ab}\bar{\sigy}\right)
\,.
\eea
where we have employed a useful constraint equation that may be
constructed from the spatial trace of the 5D Einstein tensor,
Eq.~(\ref{5DXL}), using Eqs.~(\ref{5D00}) and (\ref{5D44}),
\be
\label{yconstraint}
\bar\A''+ 2\frac{n'}{n}\bar\A' 
+ 3\left( \bar\R''+2\frac{a'}{a}\bar\R'\right)
+\nabla^2\bar\sigy'
+2\frac{a'}{a}\nabla^2\bar\sigy=0 \,.
\ee

The perturbed 5D Einstein equations, $^{(5)}\delta G^\mu_\nu=0$, in
the brane Gaussian normal gauge [see Eqs.~(\ref{5D00}), (\ref{5D0i}),
(\ref{5DXL}) and (\ref{5dtrace})] can thus be written in terms of the
perturbed 4D Einstein tensor, given in Eqs.~(\ref{4D00})-(\ref{4Dij})
and the Weyl-fluid perturbations, above, as
\bea
\label{proEin00}
{}^{(5)} \delta G^0_{~0}&=&{}^{(4)} \delta G^0_{~0}
+\kappa^2_4 \widetilde{\delta\rho} 
+2\frac{\ab'}{\ab}\left(
\nabla^2\bar{\sigy} +3\bar{\R}'\right) = 0 \,, \\
\label{proEin0i}
{}^{(5)} \delta G^0_{~i}&=&{}^{(4)} \delta G^0_{~i}
- \kappa^2_4 \widetilde{\delta p}_{,i}
+\frac{\ab^2}{\nb^2}\frac{\ab'}{\ab}\bar{\B}_{,i} = 0 \,, \\
\label{proEinX}
\Xfive &=& \Xfour 
- \kappa^2_4 \widetilde{\delta P}
+\frac{2}{3}
\left(\frac{\nb'}{\nb}+\frac{\ab'}{\ab}\right)\left(\nabla^2\bar\sigy
+3\bar\R' \right) + 2\frac{\ab'}{\ab}\bar\A'  = 0 \, , \\
\label{proEinY}
\Yfive &=& \Yfour 
- \kappa^2_4 \widetilde{\delta\pi}
-\left(\frac{\ab'}{\ab}+\frac{\nb'}{\nb}\right)\bar{\sigy} = 0 \, .
\eea

These equations, (\ref{proEin00})-(\ref{proEinY}), can be finally
rewritten, using the junction conditions~(\ref{a'/a}) and
(\ref{juncA})-(\ref{juncS}), to give the perturbed field
equations on the brane, as the linear perturbation of
Eq.~(\ref{modEinstein}),
%
%
%
%
\be
\label{pertmodEinstein}
{}^{(5)}\delta G^{\mu}_{\nu}={}^{(4)} \delta G^{\mu}_{\nu}
+\delta E^{\mu}_{\nu} - \kappa^2_4 \delta T^{\mu}_{\nu}
-\kappa^4_5 \delta \Pi^{\mu}_{\nu} = 0 \, ,
\ee
where $\delta T^{\mu}_{\nu}$ is the perturbation of the conventional
matter energy-momentum tensor given in
Eqs.~(\ref{dT00}--\ref{deltapidecomposed}) and the perturbed quadratic
energy momentum tensor is given by
\bea
\delta \Pi^{0}_{0} &=& -\frac{1}{6}\rho\delta\rho \, , \\
\delta \Pi^{0}_{i} &=& \frac{1}{6} \; \rho \;
\delta p_{,i} \, , \\
\delta\Pi^i_j &=& \delta^i_j \delta\Pi_T + \left(\nabla^i\nabla_j -
\frac{1}{3} \nabla^2\right)\delta\Pi_{TF} \, , 
\eea
where $\delta\Pi_{T}$ and $\delta\Pi_{TF}$ are the trace and the traceless
part of the anisotropic stress tensor,
\bea
\delta\Pi_{T} &=& \frac{1}{6} \left[ (\rho+P)\delta\rho
+\rho\delta P \right] \, , \\
\delta\Pi_{TF} &=& -\frac{1}{12}\left(\rho+3P\right)
\delta\pi \,.
\eea
%

There are three additional field equations for scalar perturbations
which come from the additional $^{(5)} \delta G^0_{~4}$, $^{(5)}
\delta G^4_{~i}$ and $^{(5)} \delta G^4_{~4}$ Einstein equations in
the 5D setting. In the next section we will show how the first two of
these yield energy and momentum conservation equations for ordinary
matter on the brane. The third has no counterpart in the 4D theory,
and is most neatly expressed in the brane-GN gauge in the form given in
Eq.~(\ref{yconstraint}).

%


\subsubsection{Vector perturbations}

Using the 5D vacuum Einstein equations,
(\ref{5dvecG0i})-(\ref{5dvecG4i}), we can write the vector
contribution to the projected Weyl tensor given in equations
(\ref{vecE0i}-\ref{vecEij}) as
\begin{eqnarray}
\kappa^2_4 \widetilde{\delta p}^{\rm{(vector)}} \,\ei = - \delta E^0_{~i}
 &=&
\frac{\ab^2}{2\nb^2}\left\{ {\s}' + 
\left(3 \frac{\ab'}{\ab} -\frac{\nb'}{\nb} \right) {\s}
\right\} \; \ei \, , \\
\kappa^2_4 \widetilde{\delta\pi}^{\rm{(vector)}} \eij = - \delta E^i_{~j}
 &=& \frac{1}{2} \left\{ {\ty}' +
2\frac{\ab'}{\ab} {\ty} \right\} \eij \, .
\end{eqnarray}

The 5D Einstein tensor for the vector perturbations on the brane can be
rewritten to give
\bea
{}^{(5)} \delta G^0_{~i} &=& {}^{(4)} \delta G^0_{~i}
+\delta E^0_{~i}
- \frac{\ab^2}{\nb^2} \frac{\ab'}{\ab} \,\s\; \ei \,, \\
{}^{(5)} \delta G^i_{~j} &=& {}^{(4)} \delta G^i_{~j}
+\delta E^i_{~j}
-\frac{1}{2}\left(\frac{\nb'}{\nb}+\frac{\ab'}{\ab}\right)
\,\ty\;\eij \,.
\eea
Using the junction conditions Eqs.~(\ref{a'/a}), (\ref{matvecK1}) and
(\ref{matvecK2}) the 5D Einstein equations can thus be rewritten as
\bea
{}^{(5)} \delta G^0_{~i} &=& {}^{(4)} \delta G^0_{~i}
+\delta E^0_{~i} -\kappa^2_4 \dpvec \ei
-\frac{1}{6}\kappa^4_5\,\rho \, \dpvec \,\ei \,, \\
{}^{(5)} \delta G^i_{~j} &=& {}^{(4)} \delta G^i_{~j} +\delta
E^i_{~j} -\kappa^2_4 \dpivec\,\eij
+\frac{1}{12}\kappa^4_5\,\left(\rho+3P\right)\,\dpivec \,\eij \,,
\eea
where the final terms in the expression for $^{(5)}\delta G^\mu_\nu$
can be identified with the vector perturbations in the quadratic
energy-momentum tensor, $\delta\Pi^\mu_\nu$, in
Eq.~(\ref{pertmodEinstein}).

\subsubsection{Tensor perturbations}

Using the 5D Einstein equation, (\ref{5Dtensor}), we can write the
tensor contribution to the perturbed Weyl-tensor given in
equation, (\ref{tensEij}), as
\be 
\widetilde{\delta\pi^i_j}^{\rm{(tensor)}} = - \delta E^i_{~j}
 = \frac{1}{2} {h}^{i~\prime\prime}_{j}
 + \frac{a_b'}{a_b} {h}^{i~\prime}_{j} \,. 
\ee
%
%

The 5D Einstein tensor for the tensor perturbations on the brane
can be rewritten as
\be {}^{(5)} \delta G^i_{~j} = {}^{(4)} \delta G^i_{~j} +\delta
E^i_{~j}
-\frac{1}{2}\left(\frac{\nb'}{\nb}+\frac{\ab'}{\ab}\right)
h^{i~\prime}_{j} \,, \ee
which can be re-expressed as the Einstein equation
\be {}^{(5)} \delta G^i_{~j} = {}^{(4)} \delta G^i_{~j} +\delta
E^i_{~j} -\kappa^2_4 \dpitens
+\frac{1}{12}\kappa^4_5\,\left(\rho+3P\right)\,\dpitens = 0
\,. 
\ee
where again the final term in the expression for $^{(5)}\delta G^i_j$
can be identified with the tensor perturbations in the quadratic
energy-momentum tensor, $\delta\Pi^\mu_\nu$, in
Eq.~(\ref{pertmodEinstein}).

\subsection{Energy-momentum conservation on the brane}


The perturbed 5D Einstein equations ${}^{(5)}\delta G^4_{~0}=0$ and
${}^{(5)}\delta G^4_{~i}=0$ in the brane-GN gauge [see
Eqs.~(\ref{5D04}) and (\ref{5D4i})] can be rewritten in terms of the
matter perturbations using the junction conditions
Eqs.~(\ref{juncA})-(\ref{juncS}) at the brane. We then get the
standard 4D local energy and momentum conservation equations for
scalar perturbations \cite{LMSW}
\bea
\label{energycons}
\dot{\delta\rho} +3\frac{\dot\ab}{\ab} \left(\delta\rho +\delta P\right)
+(\rho+P)\nabla^2\bar\sigma+\frac{\nb^2}{\ab^2}\nabla^2\delta p 
+3(\rho+P)\, \dot{\bar\R}
&=& 
\frac{12}{\kappa^4_5}\left(\frac{\dot \ab'}{\ab}- \frac{\dot\ab
    \nb'}{\ab\nb}\right) \bar{A} \,,  \\
\label{momentumcons}
\dot{\delta p}
+\left(\frac{\dot\nb}{\nb} +3\frac{\dot\ab}{\ab}\right)\delta p
+\delta P +(\rho+P) \bar{A} +\frac{2}{3}\nabla^2 \delta\pi &=& 0 \,,
\eea
where the term on the right-hand-side of Eq.~(\ref{energycons})
vanishes using the background energy-conservation Eq.~(\ref{G50}).

For the vector perturbations we get a momentum conservation equation
from the vector part of ${}^{(5)}\delta G^4_{~i}=0$,
\be
\label{vectormomentumcons}
\dot{\delta p}^{\rm{(vector)}}+\left(\frac{\dot\nb}{\nb}
+3\frac{\dot\ab}{\ab}\right) \dpvec +\nabla^2\dpivec =0 \,,
\ee
where we have used the junction conditions for vector perturbations
Eq.~(\ref{matvecK1}) and (\ref{matvecK2}).
We see that the 5D Einstein equations ensure that ordinary
energy-momentum conservation always holds on the brane in a vacuum
bulk, consistent with the Codazzi equation~(\ref{Codazzi}).


From Eq.~(\ref{modEinstein}), using the contracted Bianchi identities
($\nabla_{\mu} {}^{(4)}G^{\mu}_{\nu} =0$) and energy momentum
conservation ($\nabla_{\mu} T^{\mu}_{\nu} =0$) on the brane,
we find~\cite{SMS}
\be
\label{EmunuPimunu}
\nabla_{\mu} E^{\mu}_{\nu} = \kappa^4_5 \nabla_{\mu} \Pi^{\mu}_{\nu}
\,.  
\ee 
This may be interpreted as the production of bulk gravitons due to
high-energy effects on the brane \cite{Hebecker}. Energy-momentum is
transfered from the quadratic energy-momentum tensor,
$\Pi^{\mu}_{\nu}$, to the Weyl-fluid, $E^{\mu}_{\nu}$, but ordinary
energy-momentum, i.e., the tensor $T^\mu_\nu$, is always conserved,
even when bulk gravitons are excited.

Using the projected field equations we find conservation equations
governing the evolution of the Weyl fluid. In the background we have
from Eq.~(\ref{EmunuPimunu})
\be
\label{backWeyl}
\dot{\widetilde{\rho}}+4\frac{\dot \ab}{\ab}\widetilde{\rho} =0 \,,
\ee
where we used Eq.~(\ref{backenergy}). Thus the `dark radiation' term
in a homogeneous FRW cosmology is conserved~\cite{SMS} and its
energy-density redshifts away like ordinary radiation in an expanding
universe~\cite{BDEL,Flanagan,Gubser}.

For scalar perturbations we get an energy and an energy-momentum
conservation equation from Eq.~(\ref{EmunuPimunu}) \cite{LMSW}
\bea
\label{pertWeyl}
\dot{\widetilde{\delta\rho}}+4\frac{\dot \ab}{\ab}{\widetilde{\delta\rho}}
+\frac{\nb^2}{\ab^2}\nabla^2 \widetilde{\delta p} 
+ 4 {\widetilde{\rho}}\, \dot{\bar\R}
+{4\over3}\widetilde{\rho} \nabla^2 \bar\sigma &=& 0 \,, \nonumber \\
\dot{\widetilde{\delta p}}
+\left(\frac{\dot \nb}{\nb}+3\frac{\dot\ab}{\ab}\right)
{\widetilde{\delta p}}
+\frac{1}{3}{\widetilde{\delta\rho}}
+\frac{4}{3}\widetilde{\rho} \bar{A}
+\frac{2}{3}\nabla^2 \widetilde{\delta\pi} 
&=& \frac{\rho+P}{\lambda} \left\{
\delta\rho - 3\frac{\dot \ab}{\ab} \delta p
- \nabla^2 \delta \pi \right\} \,,
\eea
where we have used Eqs.~(\ref{energycons}) and (\ref{momentumcons}).
For the vector quantities we get a momentum conservation equation
\cite{BMW}
\be
\dot{\widetilde{\delta p}}^{\rm{(vector)}}
+\left(\frac{\dot \nb}{\nb}+3\frac{\dot \ab}{\ab}\right)
{\widetilde{\delta p}}^{\rm{(vector)}} 
+\nabla^2 \widetilde{\delta\pi}^{\rm{(vector)}} =
\frac{6(\rho+P)}{\lambda}  
\left(\frac{\dot \ab}{\ab} \dpvec+\frac{1}{2}\nabla^2 \dpivec \right) \,.
\ee

Thus the Weyl-fluid energy density is always conserved to linear order
(cf.~Eq.~(\ref{energycons})), whereas the Weyl-fluid momentum is
coupled to the comoving energy-density and anisotropic pressure of
ordinary matter. At low energies ($\rho+P\ll\lambda$) the momentum
transfer too becomes negligible, but the presence of the anisotropic
stress $\widetilde{\delta\pi}$, which appears as a free function in
the 4D equations, means that the Weyl-fluid cannot be completely
described as a conventional 4D fluid even at low energies.

\subsection{Curvature perturbations on the brane}

Local energy conservation can be used to demonstrate that the
intrinsic curvature perturbation on uniform density hypersurfaces is
conserved for adiabatic perturbations on large-scales independently of
the gravitational field equations~\cite{WMLL}.
We define the gauge-invariant quantity on the brane \cite{BST,LMSW,Heard}
\be
\zeta \equiv \bar\R+\frac{\delta\rho}{3(\rho+P)}\,.
\ee
%
%
From the perturbed energy conservation equation (\ref{energycons}) we
get an evolution equation for $\zeta$ \cite{WMLL},
\be
\dot\zeta =
- \frac{\dot \ab}{\ab} \left( \frac{\delta P - c^2_s
 \delta\rho} {\rho+P}\right) 
-\frac{1}{3}\nabla^2 \left[ \bar\sigma 
+ \frac{\nb^2}{\ab^2}\frac{\delta p}{\rho+P} \right] \,,
\ee
where $c^2_s=\dot{P}/\dot\rho$ is the adiabatic sound speed in the
background solution.
Hence, the curvature perturbation on uniform density hypersurfaces remains
constant on large scales, where the divergence of the comoving shear,
$\nabla^2[\bar\sigma+\delta p/(\rho+P)]$, can be neglected, 
if the non-adiabatic pressure perturbation vanishes, 
i.e.,
\be
\label{adiabatic}
\delta P_{\rm{nad}} \equiv \delta P - c^2_s \delta\rho = 0\,.
\ee

Using the junction conditions given above, Eqs.~(\ref{a'/a}) and
(\ref{juncR}), the curvature perturbation can be re-expressed solely in
terms of metric perturbations in the brane Gaussian normal gauge
\be
\label{GNzeta}
\zeta = \bar\R + \left( {\ab'\over \ab}-{\nb'\over \nb} \right)^{-1}
\left( \bar\R' + {1\over3} \nabla^2 \bar\sigma_y \right) \,.
\ee
The conservation of $\zeta$ for adiabatic perturbations on large
scales then follows from the $^{(5)}\delta G^0_4=0$ Einstein
equation~(\ref{5D04}). 
The adiabatic condition (\ref{adiabatic}) for matter perturbations
then corresponds to a condition on the bulk metric perturbations at
the brane 
\be
{\dot\ab \over \ab} \left( {\ab' \over \ab} - {\nb' \over \nb} \right)
\bar\A' = \left( {\dot\nb \over \nb} \right)^\prime \left( \bar\R' +
  {1\over3} \nabla^2\bar\sigma_y \right) \,.  
\ee 
Arbitrary bulk metric perturbations would not respect this
condition, but physically it will be enforced by the junction conditions
(\ref{Israel}) at the brane if the matter perturbations on the brane
are adiabatic, e.g., for a perfect fluid.

By transforming back from the brane Gaussian normal metric
perturbations in Eq.~(\ref{GNzeta}) using Eqs.~(\ref{transform})
and~(\ref{sigmatransforms}), we can write $\zeta$ in terms of the bulk
metric perturbations written in an arbitrary gauge. We have
\be
\label{arbitraryzeta}
\zeta = \R + {a'\over a}\xi
+ \left(\frac{\ab'}{\ab}-\frac{\nb'}{\nb}\right)^{-1}
\left\{ \R' -\frac{\ab'}{\ab} \Ayy +\frac{1}{\nb^2}\frac{\dot\ab}{\ab}
\left(\dot\xi+n\Ay\right)
+\left(\frac{\ab'}{\ab}\right)'\xi
+\frac{1}{3}\nabla^2\left(\sigma_y-\frac{1}{\ab^2}\xi\right)
\right\}\,.
\ee

Even if the total matter perturbation is not adiabatic, or indeed if
total energy is not conserved on the brane \cite{cvdb}, it is still
possible to define a curvature perturbation on hypersurfaces where
a given component has uniform-density, and this curvature perturbation
remains constant on large scales if this component is adiabatic and
its energy is conserved~\cite{WMLL,LMSW}.
For instance, in a Schwarzschild-Anti-de Sitter bulk
($\widetilde\rho\neq0$) the curvature perturbation on hypersurfaces of
uniform Weyl fluid effective density on the brane is~\cite{LMSW}
\be
\widetilde{\zeta} \equiv 
\bar\R+\frac{\widetilde{\delta\rho}}{4\widetilde{\rho}}
 \,,
\ee
%
%
%
which we can rewrite in terms of GN metric quantities, using the 
definition of the Weyl-fluid density~(\ref{Weyldensity})  
and its perturbation (\ref{dE00}), as
\be
\label{widetildezeta}
\widetilde{\zeta} = \bar\R 
- \frac{1}{2\nb^2} \left( 6\frac{\ab''}{\ab} + \Lambda_5 \right)^{-1}
\left(\nb^2\bar A'\right)' \,.
\ee
Using the conservation equation for the perturbed Weyl fluid 
(\ref{pertWeyl}) we get an evolution equation for $\widetilde\zeta$
\cite{LMSW},
\be
\label{dotwtz}
\dot{\widetilde{\zeta}} 
=
 -\frac{1}{3}\nabla^2 \left[ \bar\sigma 
+ \frac{\nb^2}{\ab^2}\frac{3\widetilde{\delta
  p}}{4\widetilde\rho} \right] \,.
\ee
Hence, on large scales when the divergence of the shear can be
neglected, $\widetilde\zeta$ is constant, regardless of whether the
matter perturbations are adiabatic.

In a conformally flat (Anti-de Sitter) bulk where $\widetilde\rho=0$
the Weyl-density perturbation, $\widetilde{\delta\rho}$, is
automatically gauge-invariant and Eq.~(\ref{pertWeyl}) reduces to 
\be
\dot{\widetilde{\delta\rho}}+4\frac{\dot \ab}{\ab}{\widetilde{\delta\rho}}
 = - \frac{\nb^2}{\ab^2}\nabla^2 \widetilde{\delta p} \,, 
\ee 
so that, when the divergence of the Weyl-momentum, $\nabla^2
\widetilde{\delta p}$, can be neglected on large-scales we have 
\be
\widetilde{S} \equiv \ab^4 \widetilde{\delta\rho} = {\rm constant} \,.
\ee
In a Schwarzschild-Anti-de Sitter bulk ($\widetilde\rho \neq 0$) it is the 
difference between the matter and Weyl-fluid curvature perturbations
that represents the relative Weyl entropy perturbation \cite{LMSW}
\be
\widetilde{S} \equiv \widetilde\zeta - \zeta \,.
\ee
Any Weyl-fluid density perturbation in a Anti-de-Sitter background
with $\widetilde\rho=0$ represents an effective entropy perturbation.
In conventional 4D Einstein gravity, it is sufficient to know the
curvature perturbation for ordinary matter, $\zeta$, to describe the
total gauge-invariant curvature perturbation, but in the brane-world
context it is not only ordinary matter (including its quadratic
corrections at high energy) but also the Weyl-fluid perturbation,
$\widetilde{S}$, that shapes the 4D geometry.

For adiabatic matter perturbations, $\delta P_{\rm{nad}}=0$, we expect
both $\zeta$ and $\widetilde{S}$ to remain constant on large scales, but
a relative entropy perturbation can cause a change in the total curvature
perturbation on the brane
\be
\zeta_{\rm tot} = \zeta + \widetilde{w}\widetilde{S} \,,
\ee
where the weight given to the Weyl entropy perturbation is time-dependent
and given by
\be
\widetilde{w} =
\left\{ 
\begin{array}{lc}
\left[ 3\ab^4(1+\rho/\lambda)(\rho+P) \right]^{-1}
 & {\rm for}\ \widetilde\rho=0 \\
4\widetilde\rho \left[ 3(1+\rho/\lambda)(\rho+P)+4\widetilde\rho
 \right]^{-1}
 & {\rm for}\ \widetilde\rho\neq0
\end{array}
\right. \,.
\ee
It is this total curvature perturbation $\zeta_{\rm tot}$ which is
related to other definitions of the intrinsic metric perturbation on
the brane such as the conformal Newtonian potential, $\Phi$ defined in
Eq.~(\ref{4Dlongpert}), and hence to observational data~\cite{LMSW}.


\section{Weyl tensor in the bulk}
\label{SWeyl}

In the preceding sections we have seen that the bulk metric
perturbations are felt on the brane only through the projected Weyl
tensor, $E_{\mu\nu}$, which can be thought of as an effective
energy-momentum tensor appearing on the right-hand-side of the
effective Einstein equations on the brane, Eq.~(\ref{modEinstein}).
Although the brane-observer is only influenced by the Weyl tensor at
the brane, it can be defined throughout the bulk. Any background
metric of the form introduced in Eq.~(\ref{backmetric}) necessarily
defines a projected Weyl tensor, $E_{\mu\nu}$, on any constant-$y$
hypersurface in the 5D bulk, and these 4D spacetimes are then sliced
into maximally symmetric 3-spaces with homogeneous Weyl-fluid density,
$\widetilde\rho(\eta,y)$.  Although this was defined in
Eq.~(\ref{Weyldensity}) in terms of the Weyl-fluid density on the
brane, this definition extends into the bulk and with the brane-world
density simply corresponding to $\widetilde\rho(\eta,0)$. The 5D
Einstein equations yield the evolution equations
\bea
\dot{\widetilde\rho} + 4 {\dot{a}\over a} {\widetilde\rho} &=& 0
\,,\nonumber\\
\widetilde\rho' + 4 {a'\over a} {\widetilde\rho} &=& 0 \,,
\eea
which can be integrated to give $a^4\widetilde\rho=$constant 
throughout the bulk.

Similarly the perturbed tensor $\delta{E}^\mu_\nu$ is defined
throughout the bulk, and can be decomposed, as in
Eqs.~(\ref{dE00}--\ref{dET}), into the perturbed Weyl-fluid density,
momentum, and anisotropic pressure
\bea \kappa_4^2 \widetilde{\delta\rho} &=& \left\{
  A''+\frac{n'}{n}\left(2A'-\Ayy'\right)-2\frac{n''}{n}\Ayy
\right. \nonumber \\
&& \qquad \left.  -
  \frac{1}{n^2}\left(\ddot\Ayy-\frac{\dot{n}}{n}\dot\Ayy\right)
  +\frac{1}{n}\left[\dot\Ay'+\frac{n'}{n}\dot\Ay
    +\left(\frac{\dot{n}'}{n}-\frac{\dot{n}n'}{n^2}\right)\Ay
  \right] \right\} \,,\nonumber\\
\kappa_4^2 \widetilde{\delta p} &=& - \frac{1}{2}\frac{a^2}{n^2}
\left\{
\B'+\left(3\frac{a'}{a}-\frac{n'}{n}\right)\B 
+\frac{2}{a^2}\left(\dot\Ayy-\frac{\dot a}{a}\Ayy\right)
+\frac{n}{a^2}\left[\left(\frac{a'}{a}-2\frac{n'}{n}\right)\Ay
-\Ay'\right]
 \right\}\,,
 \nonumber\\
\kappa_4^2 \widetilde{\delta\pi} &=& \frac{1}{a^2}\left(a^2\sigma_y\right)'
+\frac{1}{a^2}\Ayy \,, 
\eea 
where the isotropic pressure perturbation is $\widetilde{\delta
  P}=\widetilde{\delta\rho}/3$. 
In any Gaussian normal gauge these reduce to the simple definitions
given in Eqs.~(\ref{Weylfluid}) at the brane.

The energy and momentum conservation equations (\ref{pertWeyl}) in an
arbitrary gauge in the bulk can be written as
\bea
\label{bulkpertWeyl}
\dot{\widetilde{\delta\rho}}+4\frac{\dot a}{a}{\widetilde{\delta\rho}}
+\frac{n^2}{a^2}\nabla^2 \widetilde{\delta p} + 4 {\widetilde{\rho}}\, \dot{\R}
+{4\over3}\widetilde{\rho} \nabla^2 \sigma &=& 0 \,, \nonumber \\
\dot{\widetilde{\delta p}}
+\left(\frac{\dot n}{n}+3\frac{\dot a}{a}\right)
{\widetilde{\delta p}}
+\frac{1}{3}{\widetilde{\delta\rho}}
+\frac{4}{3}\widetilde{\rho} {A}
+\frac{2}{3}\nabla^2 \widetilde{\delta\pi} 
&=& 
{1\over3} \left( {a'\over a} \right)^{-1}
 \left( {a'\over a} - {n'\over n} \right) 
 \left\{
\widetilde{\delta\rho}
 - 3\frac{\dot a}{a} \widetilde{\delta p}
 + {2 \over \kappa_4^2 a^2}\nabla^2 \lR 
 \right\} \,,
\eea
where $\lR$ is the curvature perturbation in the 5D longitudinal
gauge, defined in Eq.~(\ref{5Dlongpert}). For a separable background
bulk the energy and momentum conservation equations
(\ref{bulkpertWeyl}) reduce to the standard 4D energy and momentum
conservation equations (\ref{energycons}) and (\ref{momentumcons})
with the 5D effect being due to the anisotropic stress
$\widetilde{\delta\pi}$.

In an Anti-de Sitter bulk ($\widetilde\rho=0$) the perturbed
Weyl-fluid density, momentum, and anisotropic pressure are all
naturally gauge-invariant. However in a Schwarzschild-Anti-de
Sitter bulk ($\widetilde\rho\neq0$) the perturbed density and momentum
transform under general gauge transformations (\ref{shift}) as
\bea
\widetilde{\delta\rho} &\to& \widetilde{\delta\rho} + 4
\widetilde\rho \left( {\dot{a}\over
  a} \, \delta\eta\, +\, {a'\over a} \, \delta y \right)
\,,\nonumber\\
\widetilde{\delta p} &\to& \widetilde{\delta p} +
{4\over3}{\widetilde\rho} \, \delta\eta  \,.
\eea
The anisotropic stress, $\widetilde{\delta\pi}$, remains gauge-invariant.

Because the density perturbation has the same gauge-transformation
properties as the intrinsic curvature perturbation, $\R$ in
Eq.~(\ref{transform}), it is natural to construct the
gauge-invariant Weyl-fluid density perturbation as
\be
\widetilde{\delta\rho} + 4\widetilde\rho \R
 = 4\widetilde\rho \widetilde\zeta \,,
\ee
where $\widetilde\zeta$ is the gauge-invariant curvature perturbation
on uniform Weyl-fluid density hypersurfaces defined in
Eq.~(\ref{widetildezeta}).  We find that $\widetilde{\zeta}$ is a 5D
gauge-invariant quantity constructed in terms of the metric
perturbations on any 3D spatial hypersurface at fixed $y$ and $\eta$.
In particular, it is independent of the brane location, $\xi$, unlike
$\zeta$ in Eq.~(\ref{arbitraryzeta}) which is defined with respect to
ordinary matter on the brane.

Although our original motivation was to construct a conserved quantity
on the brane, we see that $\widetilde{\zeta}$ can be evaluated on any
fixed $y$ and $\eta$ hypersurface, i.e., in any bulk gauge, and off
the brane.
Just as one can solve for the behaviour of $\widetilde\rho(\eta,y)$ in
the whole bulk due to the assumption of 3D-spatial homogeneity, so we
see from Eq.~(\ref{dotwtz}) that $\widetilde\zeta$ should be constant
throughout the bulk when the spatial gradients which appear on the
right-hand-side of Eq.~(\ref{dotwtz}) can be neglected. In this case
it may be possible to extend the picture developed in conventional 4D
cosmology to follow the evolution of large-scale perturbations as
essentially separate FRW universe, to model the evolution in the bulk
as being due to separate FRW slices of a 5D bulk.
The virtue of having a gauge-invariant definition of the Weyl-density
perturbation on spatially flat hypersurfaces, is that it can be
calculated from bulk metric perturbations in other gauges, such as the
5D-longitudinal gauge master variable, $\Omega$ in
Eq.~(\ref{scalarmastereom}), or the transverse-tracefree GN gauge
$\bar{A}$ in Eq.~(\ref{Aeom}).

The Weyl-fluid momentum is independent of the bulk
gauge, $\delta y$, but transforms in an Anti-de Sitter bulk under
shifts in the temporal gauge, like the 3D shear perturbation, $\sigma$
in Eq.~(\ref{transform}). Thus a simple gauge-invariant expression for
the Weyl-fluid momentum perturbation in the bulk is
$\widetilde{\delta p} + 4a^2\widetilde\rho \sigma/(3n^2)$,
which is the Weyl fluid momentum perturbation in the 5D longitudinal
gauge. In conventional 4D Einstein gravity it is common to choose a
temporal gauge with vanishing 3-momentum, $\delta p$, on which one can
can construct gauge-invariant definitions of the comoving density or
curvature perturbation~\cite{KS}. But in the 5D bulk the density or
curvature perturbations comoving with the Weyl-fluid momentum retain a
residual gauge-dependence due to the different possible bulk
slicings. In this sense the curvature perturbation on uniform
Weyl-fluid density hypersurfaces (or, equivalently, the Weyl fluid
density perturbation on uniform curvature hypersurfaces) is more
fundamental in the 5D case.

\section{Summary and conclusion}
\label{Ssummary}

In this paper we have presented the governing equations for
cosmological perturbations starting in Gaussian normal background
coordinates, in which the brane is at fixed coordinate $y=0$, but
allowing arbitrary linear perturbations both of the bulk metric and of
the matter perturbations on the brane.  In order to make clear the
comparison with the conventional coordinate-based approach in
four-dimensional gravity \cite{Bardeen}, we have decomposed the
perturbations into scalar, vector and tensor perturbations on the flat
three-dimensional spatial hypersurfaces.  Scalars, vectors and tensors
then obey decoupled evolution and constraint equations that we have
given in (3D) spatially gauge-invariant variables, but for arbitrary
temporal and bulk gauges.  We have then defined gauge-invariant
perturbations that correspond to different possible gauge choices
which have been employed to study bulk and brane perturbations.

One approach is to use a transverse and tracefree gauge in a Gaussian
normal coordinate system \cite{RS2,GS,HHR,Sasaki}. This is
particularly well-suited to the study of the free gravitational field
on a de Sitter slicing where the bulk wave equation is separable and
solutions can be written down in terms of harmonic functions on the
maximally-symmetric 4D spacetime. However we have shown that for any
other background cosmology the transverse-tracefree conditions in a GN
gauge over-constrain the perturbations leaving no non-vanishing scalar
perturbations.

An alternative approach to study the free gravitational field in the
bulk has been developed by Mukohyama \cite{Mukoh} and others
\cite{Kodama} working in a 5D generalisation of the longitudinal gauge
\cite{cvdb,Veneziano}. Gauge-invariant scalar, vector and tensor
perturbations can be written in terms of three master variables which
obey simple 5D wave equations in the bulk.  Separability of the 5D
equations has restricted analytic solutions to the case of a de Sitter
brane-cosmology \cite{LMW,BMW}.  In the 5D longitudinal gauge, once
the gauge-freedom in the bulk is eliminated, the perturbed brane
location, $\tilde\xi(x^\mu)$, becomes a gauge-invariant 4D
perturbation and this is not determined by the bulk field equations,
but rather by the anisotropic stress of the matter on the brane.

Matter perturbations on the brane are coupled to the bulk metric
perturbations through the Israel junction conditions at the brane.
These junction conditions only have a simple form in a Gaussian normal
coordinate system where the brane location is held fixed at $y=0$,
which we refer to as a brane GN gauge. Metric perturbations in the
brane-GN gauge coincide with the induced metric perturbations seen by
the brane bound observer. We have shown how the brane-world observer
sees the effect of the bulk metric perturbations only through the
contribution of the projected Weyl tensor to the induced Einstein
equations. This `Weyl-fluid' has an effective energy density,
momentum and pressure given in terms of second derivatives with
respect to the normal coordinate of the GN metric perturbations
\cite{Langlois}. It is the Weyl-fluid perturbations that offer
distinctive signatures of the brane-world scenario.

The four-dimensional brane observer's perspective enables us to
establish two important results for the evolution of perturbations on
large scales. Firstly energy-conservation on the brane ensures that
the curvature perturbation on uniform-density hypersurfaces remains
constant for adiabatic matter perturbations when gradient terms are
negligible \cite{WMLL}. Secondly, an analogous curvature perturbation
on hypersurfaces of uniform Weyl-fluid density on the brane also
remains constant on large scales \cite{LMSW}. We have given
gauge-invariant definitions of these quantities in terms of bulk
metric perturbations. While the matter perturbations are necessarily
defined only on specific bulk slicings (i.e., the brane or some
arbitrary extension into the bulk) we can define Weyl-fluid
perturbations that are gauge-invariant throughout the bulk allowing
the Weyl tensor evolution to be described in terms of a local density,
momentum and pressures.

Our partial understanding of brane-world metric perturbations thus far
allows us to predict the initial amplitude of perturbations in
idealised (de Sitter) models of inflation
\cite{GS,HHR,LMW,BMW,Sasaki}. For perturbations that remain constant
on super-horizon scales we can estimate their initial amplitude before
horizon re-entry \cite{LMW,BMW,LMSW,Heard}. This amounts to calculating
brane-world corrections to conventional 4D results. But we have yet to
calculate the amplitude of any intrinsically 5D effects, such as the
interplay between bulk and brane metric perturbations at horizon entry
in the radiation dominated era. This is where novel effects might
appear, even at low energies, which could be testable in astronomical
surveys of cosmological backgrounds. The outstanding challenge is to
make detailed predictions, probably requiring numerical solutions, of
the evolution of the coupled matter and metric perturbations on
cosmological time- and length-scales.

\acknowledgments

The authors would like to thank Cedric Deffayet, Stefano Foffa, 
Andrei Frolov, Lev Kofman, David Langlois, Roy Maartens and 
Misao Sasaki for useful discussions.
HAB is supported by the EPSRC, DW by the Royal Society.
Algebraic computations of tensor components were performed using
the GRTensorII package for Maple.

%
%


\appendix

\section{The background Einstein and Weyl tensors}

The 5D Einstein tensor in the background given
by the line element Eq.~(\ref{backmetric}) is
\begin{eqnarray}
\label{G00}
^{(5)} G^0_{~0} &=& 3 \left\{
\frac{a''}{a}+\left(\frac{a'}{a}\right)^2 -\left(\frac{\dot
a}{an}\right)^2 \right\} \, , \\
\label{G0i}
^{(5)} G^0_{~i} &=& 0 \, , \\
\label{G50}
^{(5)} G^4_{~0} &=& 3 \left(\frac{\dot a n'}{an}
-\frac{\dot a'}{a}\right) \, , \\
\label{G55}
^{(5)} G^4_{~4} &=&
3\left\{\frac{a'n'}{an}+\left(\frac{a'}{a}\right)^2
-\left(\frac{\dot a}{an}\right)^2 -\frac{\ddot a}{a n^2}
+\frac{\dot a \dot n}{an^3} \right\} \, , \\
\label{Gij}
^{(5)} G^i_{~j} &=&
\left\{\frac{n''}{n}+2\frac{a''}{a}+2\frac{a'n'}{an}
+\left(\frac{a'}{a}\right)^2 -\frac{1}{n^2}\left[ 2\frac{\ddot
a}{a} -2\frac{\dot a \dot n}{an} +\left(\frac{\dot a}{a}\right)^2
\right] \right\} \delta^i_{~j} \, ,
\end{eqnarray}
and the background Weyl tensor is given by
\begin{eqnarray}
\label{E00}
E^0_{~0} &=& \frac{1}{2} \left\{ \frac{a''}{a} -\frac{n''}{n}
-\left(\frac{a'}{a}\right)^2 +\frac{a'n'}{an} +\frac{1}{n^2}
\left[ \left( \frac{\dot a}{a} \right)^2
+\frac{\dot a \dot n}{an} -\frac{\ddot a}{a} \right] \right\} \, , \\
\label{Eij}
E^i_{~j} &=& -\frac{1}{3} E^0_{~0} \delta^i_{~j} \, .
\end{eqnarray}
The 4D Einstein tensor in the background is
\begin{eqnarray}
\label{4DG00}
^{(4)} G^0_{~0} &=& -3 
\left(\frac{\dot a}{an}\right)^2  \, , \\
\label{4DGij}
^{(4)} G^i_{~j} &=&
-\frac{1}{n^2}\left[ 2\frac{\ddot
a}{a} -2\frac{\dot a \dot n}{an} +\left(\frac{\dot a}{a}\right)^2
\right] \delta^i_{~j} \, .
\end{eqnarray}

\section{The perturbed Einstein tensor}

\subsection{Perturbed 4D-Einstein tensor}

We reproduce here the form of the perturbed 4-D Einstein tensor on
the brane.

\subsubsection{Scalar perturbations}

\bea
\label{4D00}
{}^{(4)}\delta  G^0_{~0} &=& \frac{2}{\nb^2}\left[
3\left(\frac{\dot\ab}{\ab}\right)
\left(-\dot{\R}+\left(\frac{\dot\ab}{\ab}\right){\A}\right)
-\left(\frac{\dot \ab}{\ab}\right)\nabla^2{\sigma}
+\left(\frac{\nb}{\ab}\right)^2\nabla^2{\R} \right] \, , \\
\label{4D0i}
{}^{(4)}\delta G^0_{~i} &=& -\frac{2}{\nb^2}\left[ \left(\frac{\dot
\ab}{\ab}\right) {\A}- \dot{\R}
\right]_{,i} \, , \\
\label{4Dij}
{}^{(4)}\delta G^i_{~j} &=& \frac{2}{\nb^2}\left[
\left\{2\frac{\ddot\ab}{\ab}+\left(\frac{\dot\ab}{\ab}\right)^2
-2\frac{\dot\ab\dot\nb}{\ab\nb}\right\} {\A} +
\left(\frac{\dot \ab}{\ab}\right) \dot{\A} -\ddot{\R}
-\left(3\frac{\dot
\ab}{\ab}-\frac{\dot\nb}{\nb}\right)\dot{\R}
\right] \delta^i_{~j} \nonumber \\ &\qquad& +\frac{1}{\nb^2}
\left[ \nabla^2 \left\{
\left(\frac{\nb}{\ab}\right)^2\left({\R}+{\A}\right)
+\left(\frac{\dot\nb}{\nb}-3\frac{\dot\ab}{\ab}\right){\sigma} 
- \dot{\sigma}\right\}~\delta^i_{~j} \right.
\nonumber \\
&\qquad& \left. -\left\{
\left(\frac{\nb}{\ab}\right)^2\left({\R}+ {\A}\right)
+\left(\frac{\dot\nb}{\nb}-3\frac{\dot\ab}{\ab}\right)
{\sigma} - \dot{\sigma} \right\}^{~i}_{,~j} \right] \, ,
\eea
It is useful to split the spatial part of the perturbed 4D-Einstein
tensor ${}^{(4)}\delta {G}^i_{~j}$ into a trace and a traceless part,
\be
{}^{(4)}\delta {G}^i_{~j} \equiv \delta^i_{~j} \; \Xfour + \left(
\nabla^i \nabla_j - \frac{1}{3}\delta^i_{~j} \nabla^2 \right) \Yfour
\,, 
\ee
where $\Xfour$ is the spatial trace and the traceless part is
$\Yfour$,
\bea
\label{4DGL}
\Xfour &=&
\frac{2}{\nb^2}\left[\left(2\frac{\ddot\ab}{\ab}
+\left(\frac{\dot\ab}{\ab}\right)^2
-2\frac{\dot\ab\dot\nb}{\ab\nb}\right) {\A} + \left(\frac{\dot
\ab}{\ab}\right) \dot{{\A}} -\ddot{\R} -\left(3\frac{\dot
\ab}{\ab}-\frac{\dot\nb}{\nb}\right)\dot{\R}\right] \nonumber \\
{} && + \frac{2}{3} \left\{ \frac{1}{\ab^2} \nabla^2 \left({\R}
+{\A} \right) + \frac{1}{\nb^2} \left[
\left(\frac{\dot{\nb}}{\nb} - 3\frac{\dot{\ab}}{\ab}\right)
\nabla^2 {{\sigma}} - \nabla^2 \dot{{\sigma}} \right]
\right\} \, , \\
\label{4DGT}
\Yfour &=& -\frac{1}{\ab^2} \left({\R} +
{\A}\right) + \frac{1}{\nb^2} \left[ \dot{{\sigma}} -
\left(\frac{\dot{\nb}}{\nb} - 3\frac{\dot{\ab}}{\ab}\right)
{\sigma} \right] \,.
\eea

\subsubsection{Vector perturbations}
%
%
\bea
%
\label{4DvecG0i}
{}^{(4)}\delta G^0_{~i} &=& - \frac{ \nabla^2
{\t}}{2\nb^2} \; \ei \, , \nonumber
\\
{}^{(4)}\delta G^i_{~j} &=& \left( \frac{1}{2\nb^2} \right) \left[
\dot\t + \left( 3\frac{\dot\ab}{\ab}-\frac{\dot\nb}{\nb}\right) {\t}
\right] \; \eij \, .
\eea
%
\subsubsection{Tensor perturbations}

%
\begin{equation}
{}^{(4)}\delta G^i_{~j}
= \frac{1}{2\nb^2}\left\{ \ddot h^i_{~j}
-\left(\frac{\dot\nb}{\nb}-3\frac{\dot\ab}{\ab}\right)\dot h^i_{~j}
-\left(\frac{\nb}{\ab}\right)^2\nabla^2 h^i_{~j}
\right\} \, .
\end{equation}
%
%

\subsection{Perturbed 5D-Einstein tensor}

We present here the first-order perturbations of the five-dimensional
Einstein tensor obtained for the perturbed metric given in
Eq.~(\ref{pertmetric}).

\subsubsection{Scalar Perturbations}

The perturbed Einstein tensor for the scalar perturbations in terms of
spatially gauge invariant variables, but for arbitrary bulk and
temporal gauges, is given by
\bea
\label{5D00}
{}^{(5)} \delta G^0_{~0} &=& \frac{6}{n^2}\left[
\left(\frac{\dot a}{a}\right)^2 \A
-\frac{\dot a}{a}\dot{\R} \right]
+3\R''+ 12 \frac{a'}{a}\R'
+\frac{2}{a^2}\nabla^2\R  - \frac{2}{n^2}\frac{\dot{a}}{a}
\nabla^2 \sigma + \nabla^2 \sigma'_y + 4\frac{a'}{a}\nabla^2
\sigma_y + \frac{1}{a^2}\nabla^2\Ayy \\
\nonumber &\qquad&  - 3\left\{\frac{a'}{a}\A'_{yy}+\frac{\dot
a}{an^2}\dot\Ayy
+2\left[\left(\frac{a'}{a}\right)^2+\frac{a''}{a}\right]\Ayy
-\frac{1}{n}\left[\frac{\dot a}{a}\A'_y+ \left(2\frac{\dot a
a'}{a^2}+\frac{\dot a'}{a}\right)\Ay\right]\right\} \, , \\
\label{5D0i}
{}^{(5)} \delta G^0_{~i} &=& \left\{ \frac{2}{n^2} \left(-
\frac{\dot a}{a}\A+\dot{\R} \right) +
\frac{a^2}{2n^2}\left[\B'+\left(5\frac{a'}{a} -
\frac{n'}{n}\right)\B\right] \right. \\ \nonumber &\qquad& \left.
-\frac{1}{2n}\left[\A'_{y}+\left(2\frac{n'}{n}+\frac{a'}{a}
\right)\A_{y}\right] + \frac{1}{n^2} \left[
\dot\A_{yy}-\frac{\dot a}{a} \A_{yy} \right] \right\}_{,i}  \, , \\
\label{5D04}
{}^{(5)} \delta G^0_{~4} &=&\frac{1}{n^2}\left\{3\left[-\frac{\dot
a}{a} \A'+ 2 \left(\frac{\dot a n'}{an}-\frac{\dot a'}{a}\right)\A
+ \dot{\R}' + \frac{\dot{a}}{a} \R' +
\left(\frac{a'}{a}-\frac{n'}{n}\right)\dot{\R} \right] +
\frac{1}{2} \nabla^2 \sigma'   \right. \\ \nonumber &\qquad&
\left. + \left(\frac{a'}{a} - \frac{n'}{n}\right) \nabla^2 \sigma
+ \frac{1}{2} \nabla^2 \dot{\sigma}_y + \frac{\dot{a}}{a} \nabla^2
\sigma_y-3\frac{a'}{a}\dot\Ayy
 \right\}
+\frac{1}{n}\left[\frac{1}{2a^2}\nabla^2\Ay +
3\left(\frac{a'n'}{an}-\frac{a''}{a}\right)\Ay\right]
 \, , \\
\label{5Dij}
{}^{(5)} \delta G^i_{~j} &=& \left\{ -\frac{1}{a^2}\left(\A+\R
\right) + \frac{1}{n^2} \left[ \dot{\sigma} +
\left(3\frac{\dot{a}}{a} - \frac{\dot{n}}{n}\right) \sigma
\right] - \sigma'_{y} - \left( 3\frac{a'}{a} + \frac{n'}{n}\right)
\sigma_{y} - \frac{1}{a^2}\A_{yy} \right\}^i_{,~j} \nonumber \\ {}
&\qquad& \nonumber + \left\{ 2\left[\frac{1}{2} \A''+
\left(\frac{a'}{a}+\frac{n'}{n}\right)\A'+ \frac{1}{n^2} \left(
\frac{\dot a}{a}\dot\A +\left(2\frac{\ddot a}{a} + \frac{\dot
a^2}{a^2} -2\frac{\dot a\dot n}{an}\right)\A\right)
\right. \right.
\\ \nonumber
&\qquad&  \left. +\R''
+\left(3\frac{a'}{a}+\frac{n'}{n}\right)\R'- \frac{1}{n^2} \left(
\ddot{\R} +\left(3\frac{\dot a}{a}-\frac{\dot n}{n}\right)\dot{\R}
\right)\right]+\frac{1}{a^2}\nabla^2(\R+\A)
\\ \nonumber &\quad& -
\frac{1}{n^2} \left[ \nabla^2 \dot{\sigma}
 + \left( 3\frac{\dot{a}}{a} - \frac{\dot{n}}{n}\right) \nabla^2 \sigma
 \right] + \nabla^2 \sigma'_y + \left(3\frac{a'}{a} +
\frac{n'}{n} \right) \nabla^2 \sigma_y \\ \nonumber &\qquad&
+\frac{1}{n}\left[\dot\A'_y+\left(\frac{n'}{n}+2\frac{a'}{a}\right)
\dot\Ay+2\frac{\dot a}{a}\A'_y +\left(\frac{\dot n'}{n}-\frac{\dot
nn'}{n^2}+2\frac{\dot aa'}{a^2} +4\frac{\dot a'}{a}\right)\Ay
\right] \\ \nonumber &\qquad& -\frac{1}{n^2}\left[\ddot\Ayy-
\left(\frac{\dot n}{n}-2\frac{\dot
a}{a}\right)\dot\Ayy\right]+\frac{1}{a^2}\nabla^2
\Ayy-\left(\frac{n'}{n}+2\frac{a'}{a}\right) \A'_{yy}  \\ {}
&\qquad& \left. - 2\left(2\frac{a''}{a} + \frac{a'^2}{a^2} +
2\frac{a'n'}{an} + \frac{n''}{n}\right) \Ayy \right\}
\delta^i_{~j} \, ,
\\
\label{5D4i}
{}^{(5)} \delta G^4_{~i} &=&  \left\{ - \A'
+\left(\frac{a'}{a}-\frac{n'}{n}\right)\A -
2\R'-\frac{a^2}{2n^2}\left[ \dot{\B} + \left(5 \frac{\dot{a}}{a} -
\frac{\dot{n}}{n}\right) \B \right] \right.\\ \nonumber &\qquad&
\left. -\frac{1}{2n}\left[\dot\A_{y}+\frac{\dot
a}{a}\A_{y}\right]+ \left(\frac{n'}{n}+2\frac{a'}{a}\right)\A_{yy}
\right\}_{,i} \,, \\
\label{5D44}
{}^{(5)} \delta G^4_{~4} &=& \frac{1}{a^2}\nabla^2\A+
3\frac{a'}{a}\A'+ \frac{3}{n^2} \left\{ \frac{\dot a}{a}\dot\A
+2\left[\frac{\ddot a}{a}+\left(\frac{\dot a}{a}\right)^2
-\frac{\dot a \dot n}{an}\right]\A -\ddot{\R} +\left(\frac{\dot
n}{n}-4\frac{\dot a}{a}\right)\dot{\R} \right\} \\ \nonumber
&\qquad&  +3\left(\frac{n'}{n}+2\frac{a'}{a}\right) \R'
+\frac{2}{a^2}\nabla^2\R - \frac{1}{n^2} \left[ \nabla^2
\dot{\sigma} + \left(4\frac{\dot{a}}{a} - \frac{\dot{n}}{n}
\right) \nabla^2\sigma \right] \\ \nonumber &\qquad& +
\left(2\frac{a'}{a} + \frac{n'}{n}\right) \nabla^2 \sigma_y
+\frac{3}{n}\left[\frac{a'}{a}\dot\Ay+\left(\frac{\dot a'}{a}
+2\frac{\dot aa'}{a^2}\right) \Ay\right]
-6\left[\frac{a'n'}{an}+\left(\frac{a'}{a}\right)^2\right]\Ayy \,.
\eea
As in the 4D case, it is useful to split the spatial part of the
perturbed Einstein tensor ${}^{(5)}\delta G^i_{~j}$ into a trace and a
traceless part,
\be
{}^{(5)}\delta G^i_{~j} \equiv \delta^i_{~j} \;  \Xfive + \left(
\nabla^i \nabla_j - \frac{1}{3}\delta^i_{~j} \nabla^2 \right)
\Yfive \,, \ee
where $\Xfive\equiv \frac{1}{3} \, {}^{(5)} \delta G^k_{~k}$ is the
3-spatial trace and the traceless part is $\Yfive$.
We get
\bea
\label{5DXL}
\Xfive &=& \frac{1}{3} \left\{ \frac{2}{n^2}\left[
\left(\frac{\dot{n}}{n}-3\frac{\dot{a}}{a}\right)\nabla^2\sigma
-\nabla^2\dot{\sigma} \right] +\frac{2}{a^2}\nabla^2\left(\A+\R
\right) +2\nabla^2\sigy'
+2\left(\frac{n'}{n}+3\frac{a'}{a}\right)\nabla^2\sigy \right\}
\nonumber\\ &\qquad& +\frac{2}{n^2}\left[ \left(\frac{\dot
n}{n}-3\frac{\dot a}{a}\right)\dot\R-\ddot\R +\frac{\dot
a}{a}\dot\A +\left(2\frac{\ddot a}{a} + \frac{\dot a^2}{a^2}
-2\frac{\dot a\dot n}{an}\right)\A\right] \nonumber\\ &\qquad&
+2\left[\R''+\frac{1}{2}\A''
+\left(\frac{n'}{n}+3\frac{a'}{a}\right)\R'
+\left(\frac{a'}{a}+\frac{n'}{n}\right)\A'\right] \nonumber\\
\nonumber &\qquad&
+\frac{1}{n}\left[\dot\A'_y+\left(\frac{n'}{n}+2\frac{a'}{a}\right)
\dot\Ay+2\frac{\dot a}{a}\A'_y +\left(\frac{\dot n'}{n}-\frac{\dot
nn'}{n^2}+2\frac{\dot aa'}{a^2} +4\frac{\dot a'}{a}\right)\Ay
\right] \\ \nonumber &\qquad& -\frac{1}{n^2}\left[\ddot\Ayy-
\left(\frac{\dot n}{n}-2\frac{\dot
a}{a}\right)\dot\Ayy\right]+\frac{2}{3a^2}\nabla^2
\Ayy-\left(\frac{n'}{n}+2\frac{a'}{a}\right) \A'_{yy}  \\ {}
&\qquad&  - 2\left(2\frac{a''}{a} + \frac{a'^2}{a^2} +
2\frac{a'n'}{an} + \frac{n''}{n}\right) \Ayy \, , \\
\label{5dtrace}
\Yfive &=& \frac{1}{n^2}\left[ \dot\sigma+\left(3\frac{\dot a}{a}
-\frac{\dot n}{n}\right)\sigma\right]
-\frac{1}{a^2}\left(\A+\R+\Ayy\right)-\sigy'-
\left(\frac{n'}{n}+3\frac{a'}{a}\right)\sigy \, .
\eea
%

\subsubsection{Vector perturbations}

Using the spatially gauge-invariant vector perturbations we obtain
the perturbed Einstein tensor components,
\bea
\label{5dvecG0i}
{}^{(5)} \delta G^0_{~0} &=& 0 \, , \qquad {}^{(5)} \delta
G^4_{~4}=0 \, ,
\qquad {}^{(5)} \delta G^4_{~0}=0\\
{}^{(5)}\delta G^0_{~i} &=& \frac{1}{2 n^2}
\left\{ -\nabla^2 \t + a^2
\left[-\left(5\frac{a'}{a}-\frac{n'}{n}\right) \s-\s'
\right] \right\} \; \ei \,, \\
\label{5dvecGij}
{}^{(5)}\delta G^i_{~j} &=& -\frac{1}{2}\left\{
\left(\frac{n'}{n}+3\frac{a'}{a}\right)\ty + \ty'
+\frac{1}{n^2}\left[ \left(\frac{\dot n}{n}-3\frac{\dot
a}{a}\right)\t-\dot \t\right] \right\}\;\eij \, , \\
\label{5dvecG4i}{}^{(5)}\delta G^4_{~i} &=&
\frac{1}{2}\left\{\nabla^2 \ty +\frac{a^2}{n^2}\left[
-\left(\frac{\dot n}{n}-5\frac{\dot a}{a}\right)\s
+\dot{\s}\right] \right\} \; \ei \,. \eea
%

\subsubsection{Tensor perturbations}

The only non-zero component of the 5D Einstein tensor for the
tensor perturbations is
\begin{equation}
\label{5Dtensor}
{}^{(5)} \delta G^i_{~j} = -\frac{1}{2}\left\{
h'^i_{~j}\left(\frac{n'}{n}+3\frac{a'}{a} \right) +
\frac{1}{n^2}\left(\frac{\dot n}{n} -3\frac{\dot a}{a}\right)\dot
h^i_{~j} +\frac{1}{a^2}\nabla^2 h^i_{~j} + {h''}^i_{~j} -
\frac{1}{n^2} \ddot h^i_{~j} \right\} \, .
\end{equation}
%

\subsection{Projected Weyl tensor}

The contribution of metric perturbations in the bulk to the
modified Einstein equations on the brane in the Gaussian normal
gauge is given by the projected Weyl tensor $E_{\mu\nu}$ in
Eq.~(\ref{modEinstein}).

\subsubsection{Scalar perturbations}

For the scalar perturbations we have
\bea \label{scalE00} \delta E^0_{~0} &=&
\frac{1}{2}\left\{\frac{1}{3\ab^2}\nabla^2\bar{\A}-\bar{\A}''
+\left(\frac{\ab'}{\ab}- 2\frac{\nb'}{\nb}\right)\bar{\A}'
+\frac{1}{\nb^2}\left[\frac{\dot{\ab}}{\ab}\dot{\bar{\A}}+2\left(\frac{\ddot{\ab}}{\ab}
-\frac{\dot{\ab}\dot{\nb}}{\ab\nb}
-\left(\frac{\dot{\ab}}{\ab}\right)^2 \right)\bar{\A} \right]
\right.\nonumber \\{}& & \left. -\frac{2}{3\ab^2}\nabla^2
\bar{\R}+\bar{\R}''+ \frac{\nb'}{\nb}\bar{\R}'
+\frac{1}{\nb^2}\left[-\ddot{\bar{\R}} + \frac{\dot \nb}{\nb}
\dot{\bar{\R}} \right] \right\} + \frac{1}{6\nb^2} \left(-
\nabla^2 \dot{\bar{\sigma}}+ \frac{\dot{\nb}}{\nb} \nabla^2
\bar{\sigma} \right) + \frac{1}{6} \left(\nabla^2 \bar{\sigma}'_y
+\frac{\nb'}{\nb} \nabla^2 \bar{\sigma}_y \right) \, , \\
\label{scalE0i} \delta E^0_{~i} &=& \frac{1}{3\nb^2}\left\{
2\left(\frac{\dot{\ab}}{\ab}\bar{\A} -\dot{\bar{\R}} \right) +
\ab^2\left[\bar{\B}' + \left(2\frac{\ab'}{\ab} - \frac{\nb'}{\nb}
\right) \bar{\B} \right] \right\}_{,i} \, , \\
\label{scalEij} \delta E^i_{~j} &=& \frac{1}{3} \left\{
\frac{1}{\ab^2}\left(\bar{\A}+\bar{\R}\right)
-\frac{1}{\nb^2}\left[\dot{\bar{\sigma}}+\left(3\frac{\dot{\ab}}{\ab}
-\frac{\dot{\nb}}{\nb}\right)\bar{\sigma}\right]
-2\bar{\sigma}^{\prime}_{y}-\left(3\frac{\ab'}{\ab}-\frac{\nb'}{\nb}\right)
\bar{\sigma}_{y} \right\}^i_{,j} \nonumber
\\
{} && +\frac{1}{6} \left\{- \frac{1}{\ab^2} \nabla^2 \bar{\A} +
\bar{\A}'' + \left( 2\frac{\nb'}{\nb} - \frac{\ab'}{\ab} \right)
\bar{\A}' - \frac{1}{\nb^2} \left[ \frac{\dot{\ab}}{\ab}
\dot{\bar{\A}} - 2 \left( \frac{\dot{\ab}\dot{\nb}}{\ab\nb} -
\frac{\ddot{\ab}}{\ab} + \left(\frac{\dot{\ab}}{\ab}\right)^2
\right) \bar{\A} \right] \right. \nonumber \\ {} && \left. -
\bar{\R}'' - \frac{\nb'}{\nb} \bar{\R}' + \frac{1}{\nb^2} \left(
\ddot{\bar{\R}} - \frac{\dot{\nb}}{\nb} \dot{\bar{\R}} \right) +
\frac{1}{\nb^2} \left[ \nabla^2 \dot{\bar{\sigma}} +
\left(2\frac{\dot{\ab}}{\ab} - \frac{\dot{\nb}}{\nb}\right)
\nabla^2 \bar{\sigma}\right] +\nabla^2 \bar{\sigma}'_y + \left( 2
\frac{\ab'}{\ab} - \frac{\nb'}{\nb} \right)\nabla^2 \bar{\sigma}_y
\right\} \delta^i_j \, . \eea
As for the Einstein tensor, it is useful to split the spatial part of
the perturbed Weyl tensor into a trace and a traceless part,
\be \delta E^i_{~j} \equiv \delta^i_j \; \XE +\left(
\nabla^i\nabla_j -\frac{1}{3}\delta^i_{~j}\nabla^2 \right)\YE \,,
\ee
where $\delta E_T$ is the spatial trace and the traceless part is
$\delta E_{TF}$.
We find
\bea \label{scalEL} \XE &=& \frac{1}{18}\left\{
3\bar{\A}''-3\bar{\R}''-3\frac{\nb'}{\nb}\bar{\R}'+
3\left(2\frac{\nb'}{\nb}-\frac{\ab'}{\ab}\right)\bar{\A}'-\nabla^2\bar{\sigy}'
-\frac{\nb'}{\nb}\nabla^2\bar{\sigy}+\frac{1}{\ab^2}\nabla^2\left(
2\bar{\R}-\bar{\A}\right)   \right. \nonumber \\ &\qquad& \left.
+\frac{1}{3\nb^2}\left[3\ddot{\bar{\R}}-3\frac{\dot{\nb}}{\nb}\dot{\bar{\R}}
-3\frac{\dot\ab}{\ab}\dot{\bar{\A}}
+6\left(\frac{\dot\ab\dot\nb}{\ab\nb}+\frac{\dot\ab^2}{\ab^2}
-\frac{\ddot\ab}{\ab}\right)\bar{\A}
+\nabla^2\dot{\bar{\sigma}}-\frac{\dot\nb}{\nb}\nabla^2{\bar{\sigma}}\right]
\right\} \,, \nonumber \\
\label{scalET} \YE &=&
\frac{1}{3}\left\{\frac{1}{\ab^2}\left(\bar{\A}+\bar{\R}\right)-2\bar{\sigy}'
+\left(\frac{\nb'}{\nb}-3\frac{\ab'}{\ab}\right)\bar{\sigy}
+\frac{1}{\nb^2}\left[
\left(\frac{\dot\nb}{\nb}-3\frac{\dot\ab}{\ab}\right)\bar{\sigy}
-\dot{\bar{\sigy}}\right]\right\} \,, \eea
for the trace and the traceless part of the projected Weyl tensor,
respectively.
%

\subsubsection{Vector perturbations}

For the vector perturbations we obtain
%
%
%
\begin{eqnarray}
\label{vecE0i} \delta E^0_{~i} &=& -\frac{\ab^2}{3\nb^2}\left\{
{\s}' + \left(2 \frac{\ab'}{\ab} -\frac{\nb'}{\nb}
\right){\s} - \frac{1}{2\ab^2} \nabla^2 {\t}\right\} \;
\ei \, , \\
\label{vecEij} \delta E^i_{~j} &=& -\frac{1}{6}\left\{
\frac{1}{\nb^2} \left[ \dot{{\t}}+ \left(3\frac{\dot
\ab}{\ab} - \frac{\dot{\nb}}{\nb}\right) {\t}\right] +
2{\ty}^{\prime} +
\left(3\frac{\ab'}{\ab}-\frac{\nb'}{\nb}\right){\ty} \right\}
\; \eij \, .
\end{eqnarray}
%

\subsubsection{Tensor perturbations}
The only non-zero component is gauge-invariant and yields
\begin{equation}
\label{tensEij} \delta E^i_{~j} = \frac{1}{6}\left\{
\dot{{h}^i_{~j}}\left(\frac{\dot \nb}{\nb^3}-3\frac{\dot
\ab}{\ab\nb^2}\right) -\frac{1}{\nb^2} \ddot{{h}^i_{~j}} -2
{{h}''^i_{~j}}
+{{h}'^i_{~j}}\left(\frac{\nb'}{\nb}-3\frac{\ab'}{\ab}\right)
+\frac{1}{\ab^2}\nabla^2 {h}^i_{~j} \right\} \, .
\end{equation}
%


\end{document}